\documentclass[]{aastex631}
\usepackage{accents}
\usepackage{soul}

\def\s{\phantom}

\def\amin{\ifmmode^{\prime}\else$^{\prime}$\fi}
\def\asec{\ifmmode^{\prime\prime}\else$^{\prime\prime}$\fi}

\def\simgt{\lower.5ex\hbox{$\; \buildrel > \over \sim \;$}}
\def\simlt{\lower.5ex\hbox{$\; \buildrel < \over \sim \;$}}

\newcommand\chandra{{\it Chandra}}
\newcommand\xmm{{\it XMM-Newton}}

\newcommand\integral{{\it INTEGRAL}}
\newcommand\INTEGRAL{{\it INTEGRAL}}

\newcommand\swift{{\it Swift\/}}
\newcommand\nustar{{\it NuSTAR}}
\newcommand\suzaku{{\it Suzaku}}

\newcommand\eflux{{erg\,cm$^{-2}$\,s$^{-1}$}}

\newcommand{\mdot}{$\dot{m}$} 

\newcommand{\src}{J1745$-$3213} 

\shorttitle{Constraining WD mass of a new IP J174517.0$-$3213}
\shortauthors{Vermette et al.}

\graphicspath{{./}{figures/}}

\begin{document}

\title{\color{black} Constraining white dwarf mass and magnetic field strength of a new intermediate polar through X-ray observations}

\correspondingauthor{Ciro Salcedo}
\email{cgs2155@columbia.edu}

\author[0000-0003-4292-1281]{Benjamin Vermette}
\affiliation{Columbia Astrophysics Laboratory, Columbia University, New York, NY 10027, USA}
\author[0000-0001-8586-7233]{Ciro Salcedo}
\affiliation{Columbia Astrophysics Laboratory, Columbia University, New York, NY 10027, USA}
\author[0000-0002-9709-5389]{Kaya Mori}
\affiliation{Columbia Astrophysics Laboratory, Columbia University, New York, NY 10027, USA}
\author[0000-0001-6470-6553]{Julian Gerber}
\affiliation{Columbia Astrophysics Laboratory, Columbia University, New York, NY 10027, USA}
\author[0000-0001-8532-1273]{Kyung Duk Yoon}
\affiliation{Columbia Astrophysics Laboratory, Columbia University, New York, NY 10027, USA}
\author[0000-0002-6653-4975]{Gabriel Bridges}
\affiliation{Columbia Astrophysics Laboratory, Columbia University, New York, NY 10027, USA}
\author[0000-0002-3681-145X]{Charles J. Hailey}
\affiliation{Columbia Astrophysics Laboratory, Columbia University, New York, NY 10027, USA}

\author[0000-0002-0107-5237]{Frank Haberl}
\affiliation{Max-Planck-Institut f{\"u}r extraterrestrische Physik, Giessenbachstrasse 1, 85748 Garching, Germany}

\author[0000-0002-6089-5390]{Jaesub Hong}
\affiliation{Harvard-Smithsonian Center for Astrophysics, Harvard University, Cambridge, MA 02138, USA}
\author[0000-0002-1323-5314]{Jonathan Grindlay}
\affiliation{Harvard-Smithsonian Center for Astrophysics, Harvard University, Cambridge, MA 02138, USA}
\author[0000-0003-0293-3608]{Gabriele Ponti}
\affiliation{INAF-Osservatorio Astronomico di Brera, Via E. Bianchi 46, I-23807 Merate (LC), Italy}
\author[0000-0001-8722-9710]{Gavin Ramsay}
\affiliation{Armagh Observatory and Planetarium, College Hill, Armagh, BT61 9DG, UK}

\begin{abstract}

We report timing and broad-band spectral analysis of a Galactic X-ray source, CXOGBS J174517.0$-$321356 (J1745), with a 614-second periodicity. \chandra\ discovered the source in the direction of the Galactic Bulge. \cite{Gong2022} proposed J1745 was either an intermediate polar (IP) with a mass of $\sim$1 $M_{\odot}$, or an ultra-compact X-ray binary (UCXB). \textcolor{black}{ To confirm J1745’s nature, we jointly fit \xmm\ and \nustar\ spectra, ruling out a UCXB origin.  We have developed a physically realistic model that considers finite magnetosphere radius, X-ray absorption from the pre-shock region, and reflection from the WD surface to properly determine the IP properties, especially its WD mass. 
To assess systematic errors on the WD mass measurement, we consider a broad range of specific accretion rates ($\dot{m} = 0.6\rm{-}44$ g\,cm$^{-2}$\,s$^{-1}$) based on the uncertain source distance  ($d = 3\rm{-}8$ kpc) and fractional accretion area ($f = 0.001\rm{-}0.025$). Our model properly implements the fitted accretion column height in the X-ray reflection model and accounts for the underestimated mass accretion rate due to the (unobserved) soft X-ray blackbody and cyclotron cooling emissions. 
We found that the lowest accretion rate of $\dot{m}$ = 0.6 g\,cm$^{-2}$\,s$^{-1}$, which corresponds to the nearest source distance and maximum $f$ value, yield the WD mass of $(0.92\pm0.08) M_{\odot}$. On the other hand, as long as the accretion rate is $\dot{m} \simgt 3 $ g\,cm$^{-2}$\,s$^{-1}$, the WD mass is robustly measured to be $(0.81\pm0.06)  M_{\odot}$, nearly independent of $\dot{m}$. The derived WD mass range is consistent with the mean WD mass of nearby IPs. Assuming spin equilibrium between the WD and accretion disk, we constrained the WD magnetic field to $B \simgt 7$ MG, indicating that it could be a highly magnetized IP.
Our analysis presents the most comprehensive methodology for constraining  the WD mass and B-field of an IP by consolidating the effects of cyclotron cooling, finite magnetospheric radius, and accretion column height. 
}

\end{abstract}

\section{Introduction} \label{sec:intro} 

Cataclysmic variables (CVs) are interacting white dwarf (WD) binary systems powered by accretion via Roche-lobe overflow from a late-type main sequence companion star. 
Thanks to the extensive optical and X-ray surveys of CVs, the last 50 years have seen enormous strides in our understanding of the accretion physics, binary formation, and evolution of CVs. 
CVs are the most common type of interacting compact binaries and are also recognized as one of the candidate progenitor classes for type Ia supernovae and gravitational wave sources detectable by {\it LISA} in the mHz frequency band \citep{Hillebrandt2000, Meliani2000}. 
The majority of hard X-ray emission in our galaxy is represented by magnetic CVs (mCVs) of two sub-classes. In mCVs, infalling material is funneled onto the WD poles along magnetic field lines, heated at a stand-off shock, and emits thermal X-rays.  
Intermediate polars (IPs) are a type of mCV that have non-synchronized orbits and WD magnetic fields ($B\sim0.1-10$~MG) strong enough to truncate the inner accretion disk; they are copious emitters of hard X-rays  \citep[$kT \sim 20{\rm-}40$ keV; ][]{Mukai2017}. On the other hand, polars are mCVs with stronger WD magnetic fields ($B\sim10-240$~MG) whose interactions with companions cause the synchronization of their orbital and spin periods. Polars typically have softer X-ray spectra ($kT \sim 5\rm{-}20$ keV) than IPs due to  faster cyclotron cooling \citep{Mukai2017}. The CV population in the solar neighborhood ($d < 150$ pc) was recently determined from  an unbiased, volume-limited sample based on the {\it Gaia} optical survey, finding a surprisingly higher fraction of mCVs (36\%) than previously predicted by conventional CV formation models  \citep{Pala2020}.  

In the Galactic Center (GC), Bulge and Ridge regions at $d\sim 1-8$ kpc, the optical/UV emission is extincted by significant absorption and dust scattering, making the X-ray observations all the more important. Dedicated hard X-ray surveys by {\it RXTE}, \integral, \suzaku\ and \nustar\ revealed that the diffuse X-ray emission (which mostly represents an unresolved population of CVs) from the GC, Bulge and Ridge is distinct in each region. The \nustar\ discovery of the central hard X-ray emission (CHXE with $kT \sim 35$ keV) \citep{Perez2015}, which spatially coincides with the nuclear star cluster (NSC) in the central 10 pc  \citep{Schodel2014}, suggests that the CV population is dominated by IPs with mean WD mass $\langle M \rangle \sim 0.9M_\odot$ \citep{ Hailey2016}. Outside the CHXE/NSC region, the plasma temperatures of the Bulge and Ridge measured by \integral, \suzaku\ and \nustar\  \citep[$kT \sim8$--15~keV; ][]{Turler2010,Yuasa2012, Perez2019} are in stark contrast to the ones of the CHXE  \citep[$kT \sim 35$~keV; ][]{Perez2015}. While the CHXE is largely characterized by IPs, the diffuse X-ray emission in the Bulge and Ridge may consist of a different CV population dominated by polars or non-magnetic CVs \citep{Hailey2016,Xu2019, Perez2019}. This is also supported by an Fe line diagnostics study using \suzaku\ data of the Ridge X-ray emission \citep{Xu2016, Nobukawa2016}. The different but unresolved CV populations in these regions may have emerged due to differences in star forming history, binary evolution (e.g., more interactions with stars in the NSC), or metallicities. 
However, it is nearly impossible to deconvolve the diffuse X-ray emission into different CV populations (with varying $L_{\rm X}$ and $kT$ distributions) through analysis of diffuse hard X-ray continuum or Fe line emission only. 

The only unambiguous way to explore CV populations in the GC, Bulge and Ridge is by analyzing the detected X-ray point sources and determining their source types by analysis of their plasma temperatures or Fe emission lines. For example, in the GC region, which has extensively been observed in the X-ray band \citep{Wang2002, Muno2009}, \chandra\ ACIS analysis of X-ray point sources investigated  the CV population by confirming that low luminosity IPs ($L_{\rm X} \simlt 10^{33}$~erg\,s$^{-1}$) make up a large fraction of the CHXE \citep{Xu2019b}. However, in the Bulge region outside the GC region ($r\sim100$ pc from Sgr A*), X-ray sources and their populations are poorly understood, partially due to the lack of extensive X-ray surveys and/or follow-up of deeper X-ray observations. The situation will be improved by the ongoing \swift-XRT, \chandra/\xmm, and {\it eROSITA} X-ray surveys of the Bulge region \citep{Bahramian2021, Mondal2022}, and by follow-up observations. Indeed, follow-up \nustar\ and \chandra\ observations of unidentified {\integral} sources in the Galactic disk detected a handful of IPs \citep{Coughenour2022}. These INTEGRAL source investigations, as well as the \nustar\ legacy observations of 25 IPs \citep{Shaw2020}, have demonstrated the significant role of \nustar's broad-band X-ray spectroscopy data for identifying mCVs and their WD masses.   

Here, we report \xmm\ and \nustar\ observations of a new IP, CXOGBS J174517.0-321356 (\src\ hereafter),  located $3.2^\circ$ away from the GC. \src\ was discovered in the Galactic Bulge Survey with a high degree of X-ray absorption indicating a distance greater than 3 kpc \citep{Jonker2014}.  The archived \xmm\ observations in 2010 (originally targeting a black hole binary H1743$-$322) showed a periodicity at $P = 614$ sec \citep{Gong2020}, which was confirmed by a follow-up on-axis \xmm\ observation in 2021 \citep{Gong2022}. The  source was previously speculated to be an IP or ultra-compact X-ray binary \citep{Gong2022}. The 614-sec periodicity was interpreted as a spin period in the IP case or an orbital period in the UCXB case.  Our joint \xmm\ and \nustar\ observations in 2021 detected hard X-ray emission extending up to $\sim50$ keV. {\color{black}Combining all available \xmm\ and \nustar\ data supports the hypothesis that \src\ is an IP with WD mass  $M\sim 0.75 - 1.0 M_\odot$. }In \S2, we describe the X-ray observations and data reduction of \src. In \S3 and \S4, we present our X-ray timing and spectral analysis of \src, respectively. Based on the X-ray analysis results and the lack of an IR counterpart (\S5), we identified that \src\ is a new IP in the direction of the Bulge (\S6.1). By applying a new X-ray spectral model developed for mCVs to the \xmm\ + \nustar\ spectra, we determined the WD mass of \src\ (\S6.2). In \S7, we summarize our results and discuss the significance of constraining WD masses of IPs using broad-band X-ray spectral and timing data.

\section{X-ray observations and data reduction} 

\src\ was discovered by \chandra\ \citep{Jonker2014} and was serendipitously observed by \xmm\ in 2010 while it targeted a nearby black hole binary, H1743-322. \src\ was not in the field of view of the 2010 \xmm\ EPIC-PN detector. In the MOS data, the point source appeared elongated due to the large off-axis angle; we thus used an elliptical region with semi-major and semi-minor axes of 40\asec\ $\times$ 30\asec\ for source extraction. The background region was extracted using an elliptical annulus with an inner ring the size of the source extraction region and an outer ring with semi-major and semi-minor axes of 60\asec\ $\times$ 45\asec\, respectively.

To identify the source type, follow-up \xmm\ (observing between 0.2 - 15 keV) and \nustar\ (observing between 3 - 79 keV) observations were performed in 2021. For all of the observations, the source extraction regions were centered around  the \chandra\ position (RA = 17:45:17.0 $\pm 0.11$, DEC = -32:13:56.5 $\pm 0.11$; J2000). We reduced all \xmm\ EPIC data using SAS 19.1 and extracted source events using \emph{evselect} after applying GTI filters generated by \emph{emchain} for MOS and \emph{epchain} for PN \citep{2004ASPC..314..759G}.
The second \xmm\ observation targeted \src\ with 31 ks of exposure, and a circular region with a radius of  20\asec\ for both (MOS) and (PN) was adopted for extracting source events. The background was extracted using a circular annulus of inner and outer radii of 20\asec\ and 30\asec\, respectively.

\nustar\ observed \src\ in 2021-03 for 49 ks exposure. All data were processed using the standard \nustar\ processing procedures \citep{Harrison2013}. A $r=50$\asec\ circular region around the \chandra\ source position was used for extracting source events for both timing and spectral analysis. A circular annulus of inner and outer radii of 50\asec\ and 80\asec\ was used to extract the background. In the \nustar\ observations, the stray-light background contamination from H1743$-$322 did not overlap with our source nor with background extraction regions.

\begin{deluxetable}{lcccc}[ht]
\tablecaption{Timeline of X-Ray Observations}
\tablehead{ \colhead{Date}   
&
\colhead{ObsId}  
& 
\colhead{Target}
& 
\colhead{Telescope} 
&  
\colhead{Exposure (ks)} }
\startdata
2010-10-09  & 0553950201 & H1743$-$322 & \xmm\  & 86    \\
2021-02-28  & 0870990201 & \src\ & \xmm\ & 31   \\
2021-03-06  & 30601021002 & \src\ & \nustar\ & 49   
\enddata
In the 2010 \xmm\ observation, PN was operated in the Small Window mode putting \src\ outside of the FOV.  
\label{tab:obstimeline}
\end{deluxetable}

\section{Timing analysis} 

\begin{figure}[ht]
\centering
\begin{minipage}{.3\textwidth}
  \centering
  \includegraphics[width=6cm]{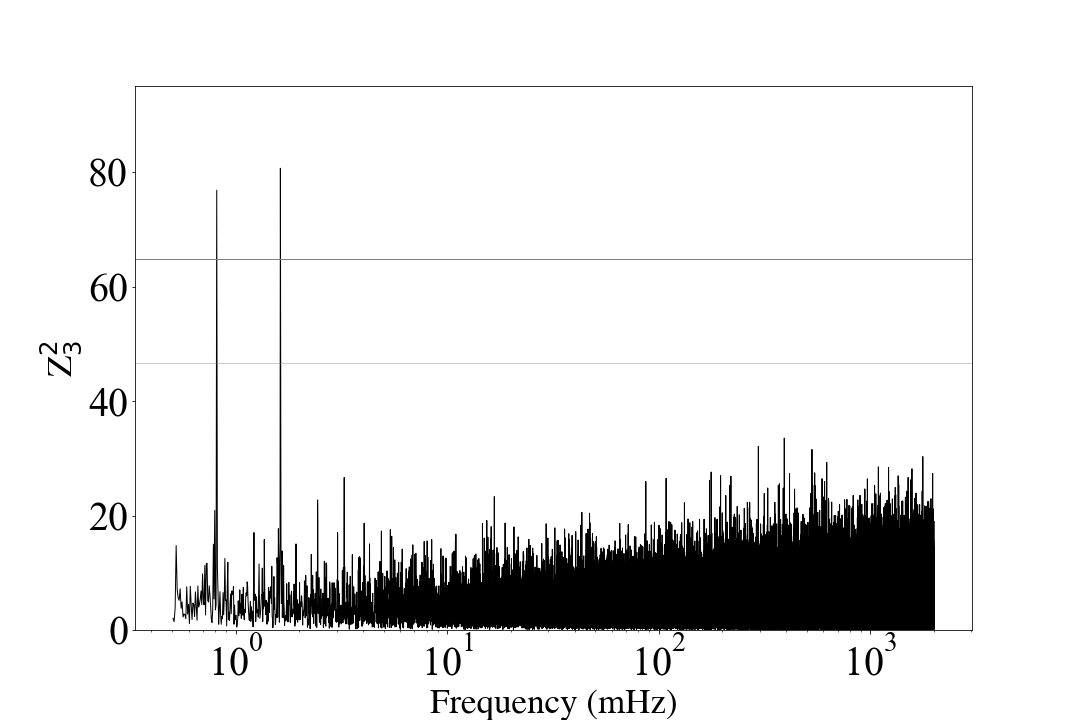}
  \label{fig:newolddMOS1broad}
\end{minipage}
\begin{minipage}{.3\textwidth}
  \centering
  \includegraphics[width=6cm]{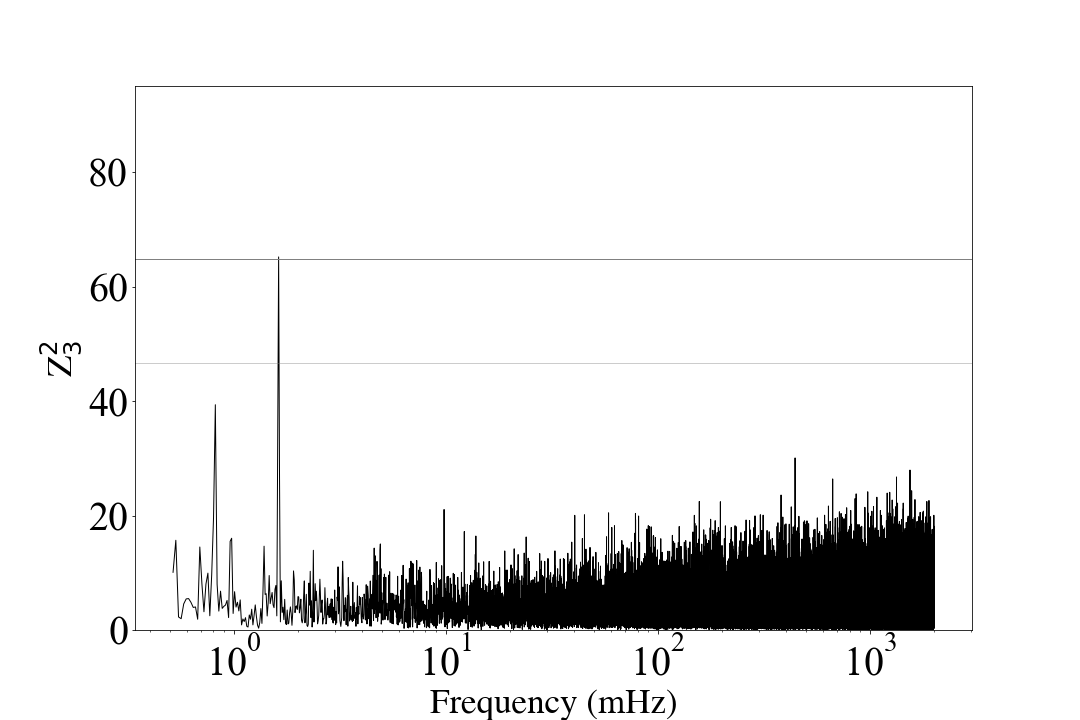}
  \label{fig:newnewMOS1broad}
\end{minipage}
\begin{minipage}{.3\textwidth}
  \centering
 \includegraphics[width=6cm]{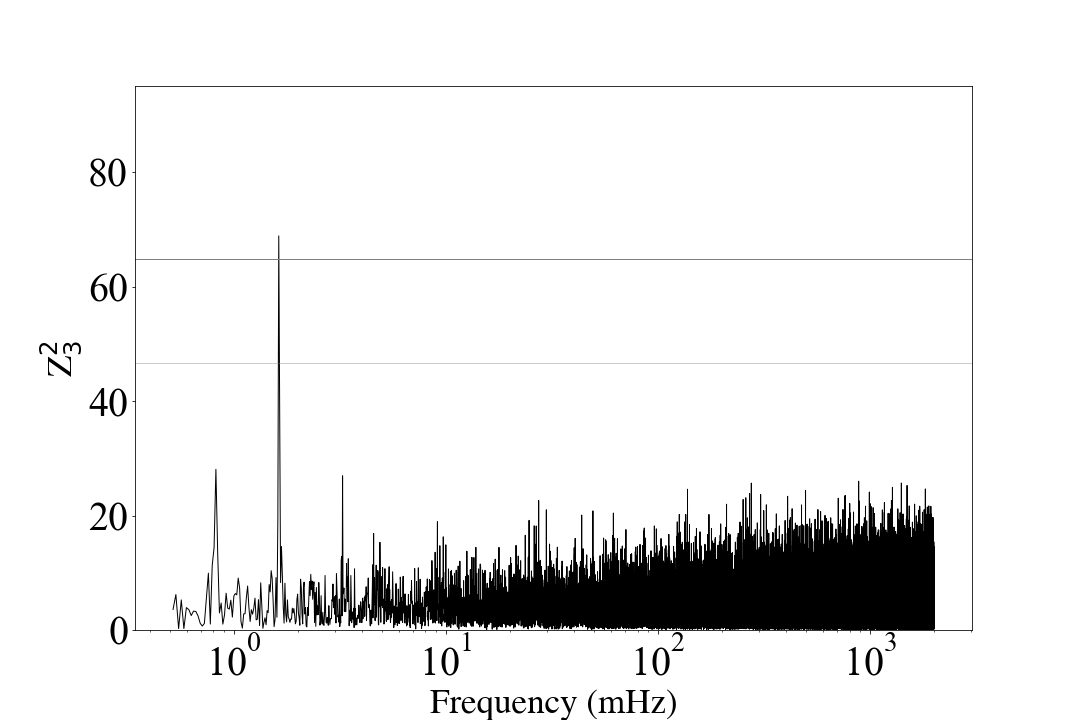}
  \label{fig:newPNbroad}
\end{minipage}%
\caption{
$Z^2_3$ broadband (0.5 mHz - 2 Hz) periodograms. From left to right: 0553950201 MOS,   0870990201 MOS and 0870990201 PN. In each periodogram, 3-$\sigma$ and 5-$\sigma$ significance levels are marked by solid and faded lines, respectively. The most significant peak is in the region of P $\sim$ 614 seconds. The first observation significantly detected a peak at 1227 s, which was less significantly detected in later observations}
\label{broadperiodograms}
\end{figure}

Here, we present our timing analysis of the \xmm\ and \nustar\ observations using the {\tt Stingray} software package \citep{matteo_bachetti_2022_6394742}.  
We applied the barycentric correction to the extracted source events using the \emph{barycen} command in SAS. The photon events were then filtered to different energy bands (0.3--10, 0.3--5 and 5--10 keV) using {\tt Hendrics} tools in {\tt Stingray}.    
For each \xmm\ observation, we analyzed MOS1 and MOS2 data jointly, while PN data (available only from the 2021 observation) was analyzed separately. 
First, we produced periodograms over a broad frequency band from $f = 5 \times 10^{-4}$ Hz to $f = 2$ Hz using {\tt Stingray}'s weighted $Z^2_n$ function \textbf{z\_n\_search} (Figure \ref{broadperiodograms}). 
In power density spectra, we found no evidence of red noise in this frequency band thus validating the use of the $Z^2$ test.  
In the soft (0.3--5 keV) band alone, we detected peaks at $f\approx 1.628$ mHz (corresponding to $P\approx614$ sec) with 3-$\sigma$ to 5-$\sigma$ significance. Furthermore we detected peaks at $f\approx 0.815$ mHz (corresponding to $P\approx1227$ sec) with less significance (especially in the newest observation where it was recorded below 3-$\sigma$ significance). We found no other periodic signals above 3-$\sigma$ level.

\begin{figure}[ht]
\centering
\begin{minipage}{.3\textwidth}
  \centering
  \includegraphics[width=6cm]{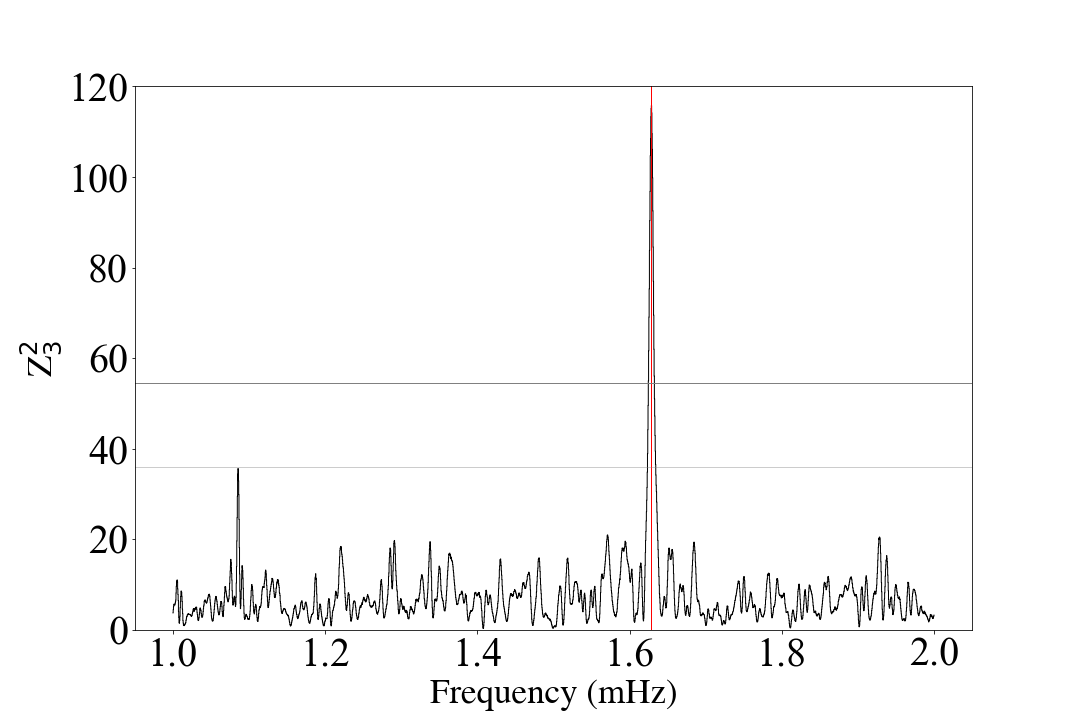}
  \label{fig:oldMOS1}
\end{minipage}
\begin{minipage}{.3\textwidth}
  \centering
  \includegraphics[width=6cm]{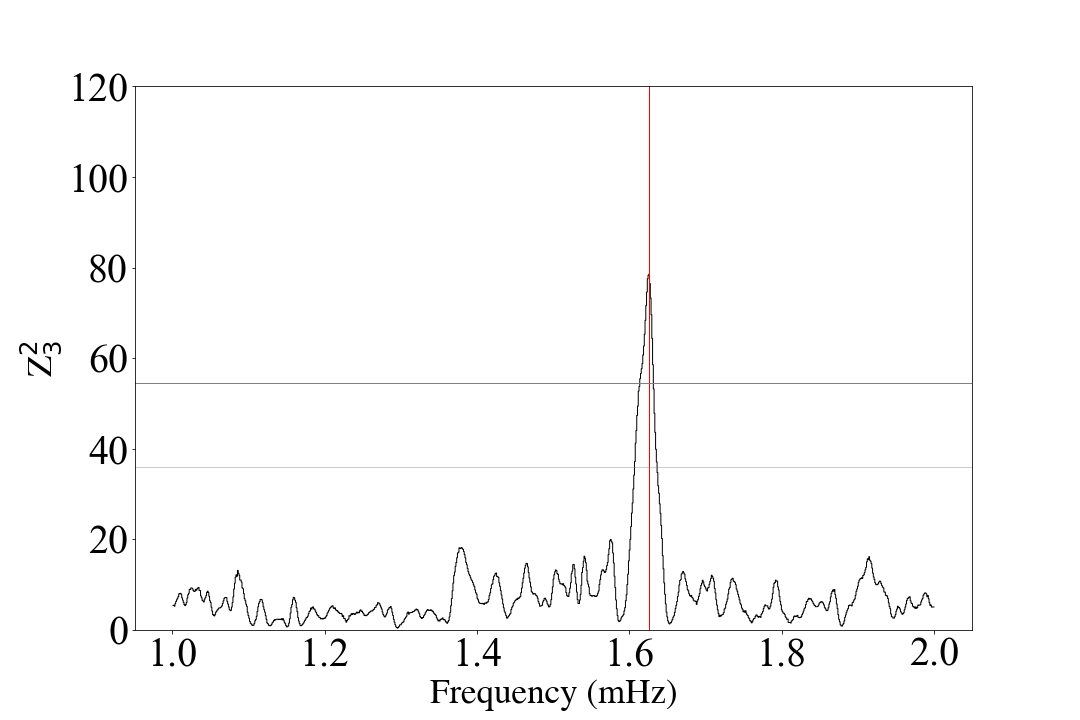}
  \label{fig:newMOS1}
\end{minipage}
\begin{minipage}{.3\textwidth}
  \centering
 \includegraphics[width=6cm]{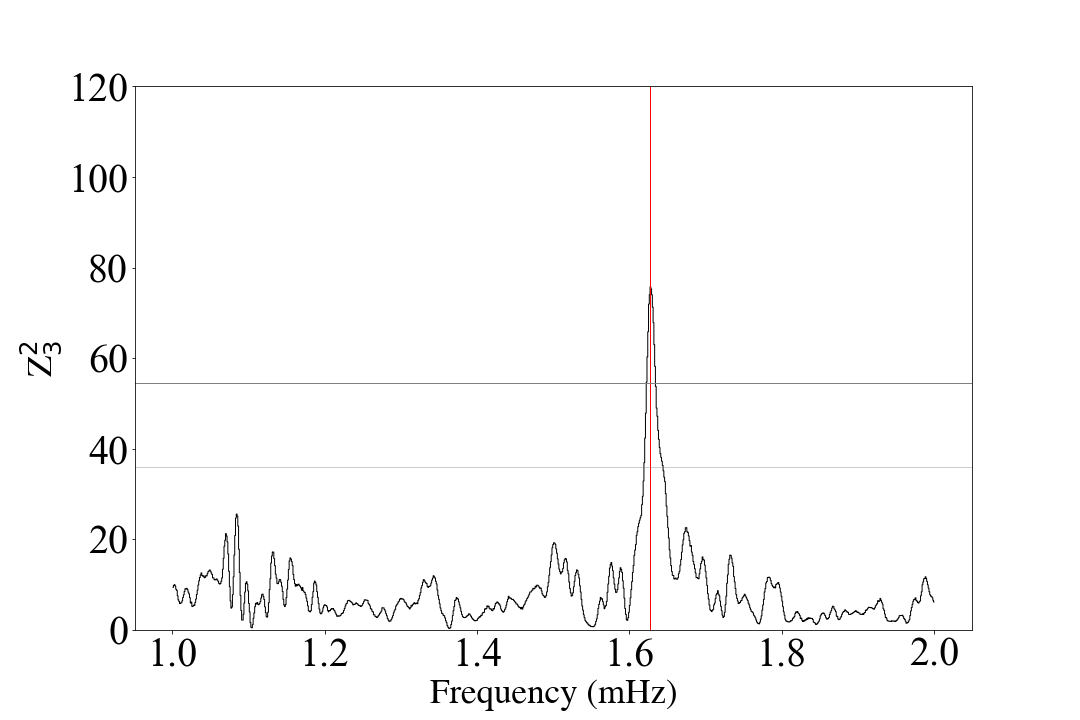}
  \label{fig:PN}
\end{minipage}%

\centering
\begin{minipage}{.3\textwidth}
  \centering
  \includegraphics[width=6cm]{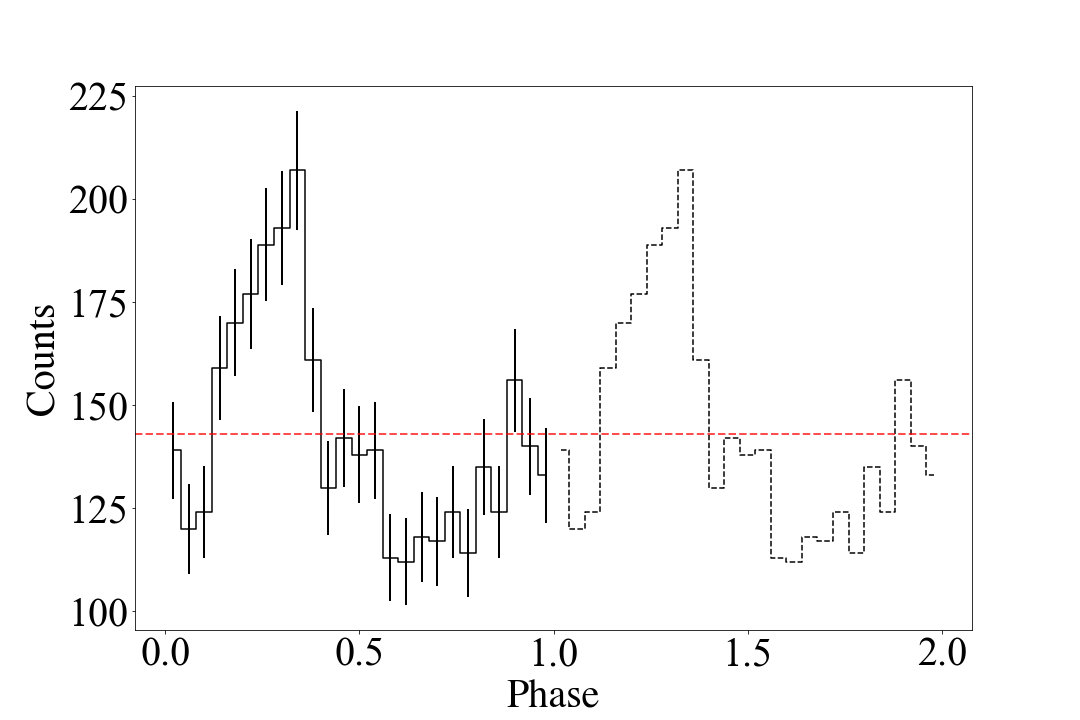}
  \label{fig:oldMOS1phase}
\end{minipage}
\begin{minipage}{.3\textwidth}
  \centering
  \includegraphics[width=6cm]{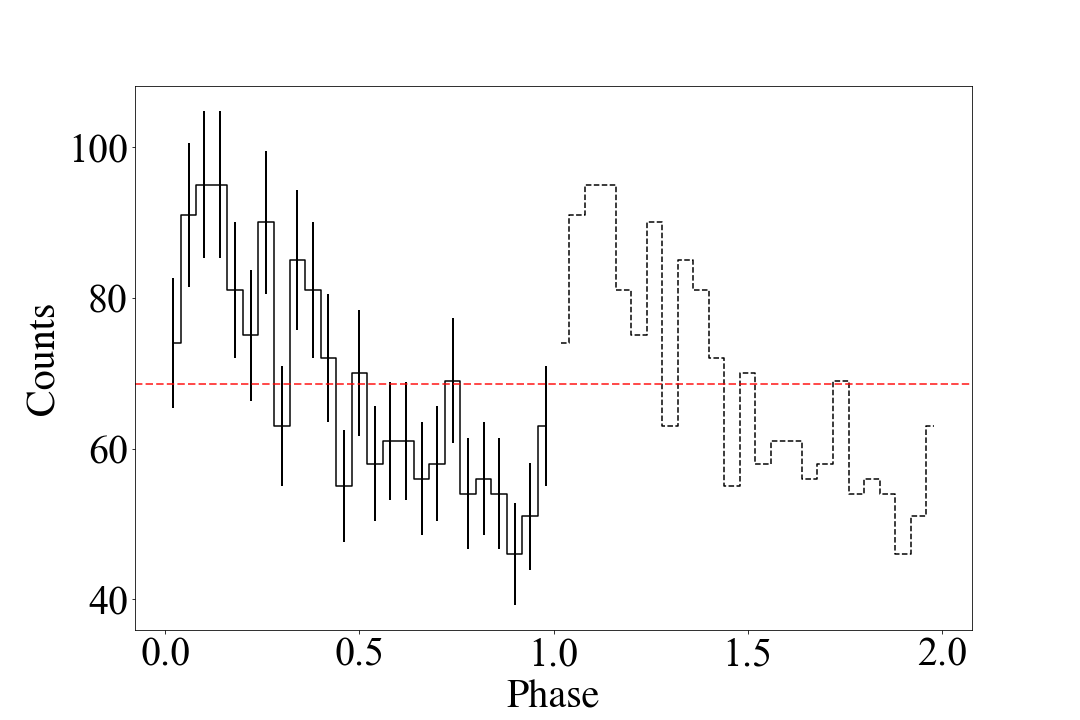}
  \label{fig:mos1newphase}
\end{minipage}
\begin{minipage}{.3\textwidth}
  \centering
 \includegraphics[width=6cm]{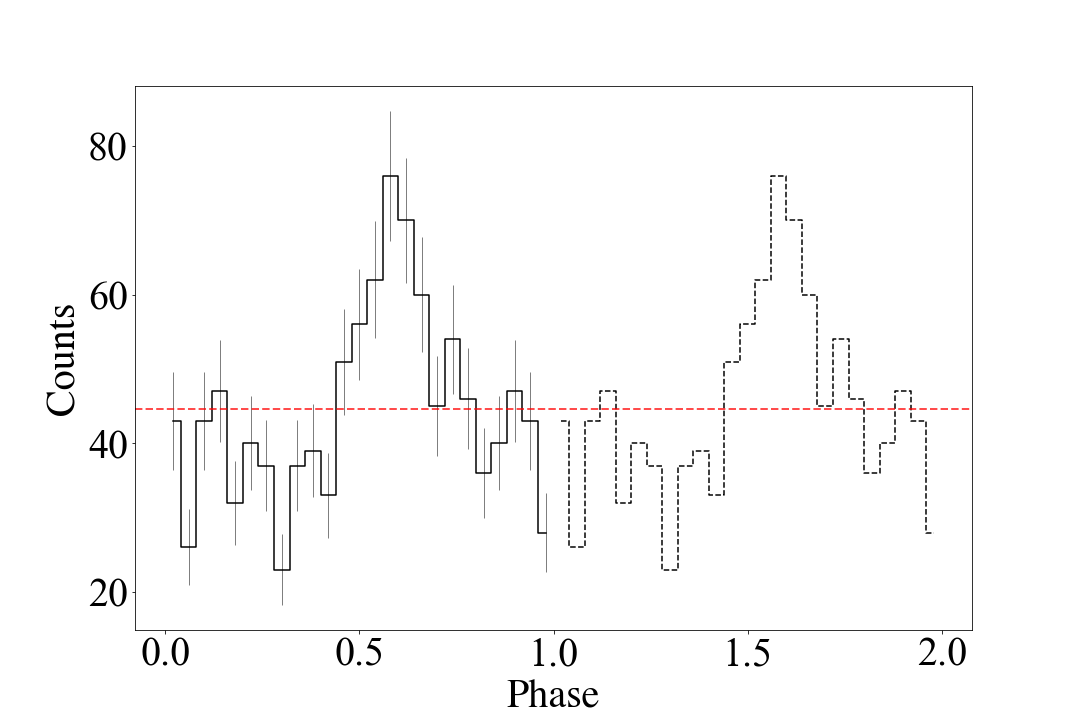}
  \label{fig:PNphase}
\end{minipage}%
  \caption{$Z^2_3$ periodograms (top) and pulse profiles folded by their peak period (bottom) and reference time at $T_0 = 55197$ MJD. From left to right: 0553950201 MOS1,   0870990201 MOS1 and 0870990201 PN. 
  In each periodogram, 3-$\sigma$ and 5-$\sigma$ significance levels are marked by solid and faded lines, respectively. In the pulse profile plots, the red horizontal lines indicate the mean counts per phase bin and 1-$\sigma$ error bars are included at  $\phi=0-1$.} 
\label{pulses}
\end{figure}

To characterize the peak at $P\sim614$ sec more accurately, we conducted $Z^2$ tests over a narrow frequency band ($f=1-2$ mHz) with a fine step size of $\Delta f = 2\times 10^{-6}$ Hz (see Figure \ref{pulses} for the periodograms in the 0.3--5 keV band). 
We found that the 614-sec peak is broad (with $\Delta f \simlt 0.003$ mHz) and, in some cases, asymmetric, possibly due to timing noise intrinsic to the source. We calculated 90\%-CL error bars on the peak periods by use of the $Z^2$ test tool in  {\tt Hendrics}  \citep{matteo_bachetti_2022_6394742}. We found that the peak periods between the two \xmm\ observations' MOS and PN data are consistent with each other within error.   

The signal was not detected in our identical analysis of background events. We also confirmed the 614-sec periodicity in the same frequency and energy bands using the epoch-folding method.
By applying the $Z^2$ test jointly on the 2010 and 2021 \xmm\ observation data, we obtained an upper limit on the period derivative to $|\dot{P}| < 4.2 \times 10^{-11}$ [s/s] (90\% CL). 

We performed similar $Z^2$ tests over a wider frequency band ($f=0.5-2$ mHz) to capture the $\sim$ 1227 s peak. We additionally saw a significant peak at $\sim$ 1841 sec. However, analysis of pulse profiles indicated that these peaks were not representative of additional physical phenomena, but rather of lower harmonics of the fundamental $\sim$ 613-sec signal. The pulsed profiles in the 0.3--5 keV band are generated by folding source events by the peak period in each \xmm\ data set. They are characterized by a single peak with $\sim30-40$\% pulsed fractions (bottom panel in Figure \ref{pulses}). The lightcurve peaks between the different observations and instruments do not appear to be aligned in phase, but we attribute this to the slightly different peak periods  (e.g., $P=615.01^{+1.13}_{-0.61}$ [sec] for MOS and $614.37^{+0.97}_{-0.95}$ [sec] for PN in the 2021 \xmm\ observation). In the 5--10 keV band where we did not detect the 614 sec period, we calculated an upper limit on the pulsed fraction using  the \textbf{HENzsearch} command to  31\%, 34\%, and 28\% for the 2010 MOS, 2021 MOS, and PN data, respectively. Combining the newest and highest quality observations (2021 MOS + PN data) yields an upper limit of modulation of 23 \% above 5 keV. The 1227 sec signal shows two symmetric non-overlapping pulses and the 1841 sec signal shows three symmetric non-overlapping pulses (See appendix for more information).Double-peaked modulation in UCXBs is characterized by asymmetric rising and falling flanks and asymmetric peaks (\cite{Annala_2010}, \cite{Norton2004}).Similarly, asymmetric peaks are also characteristic of double-peaked modulation in IPs, where the spin-folded light curve demonstrates uneven peak amplitudes \citep{10.1111/j.1365-2966.2011.20256.x}. These asymmetries would indicate that two linked physical features are being observed at a single fundamental frequency (e.g. the two poles of a rotating object). There would still be a significant signal at half of the fundamental frequency. However, to continue the example of poles, this would effectively fold the two poles together.  Again, it is easiest to check the multiples of any significant frequency to investigate whether a feature is a fundamental signal or not. Our data did not demonstrate asymmetry; the pulse profiles have peaks symmetric in both height and shape at multiples of 614 s. Thus, we maintain that the 614-sec signal represents a spin period. Notably, this stands in contrast to the conclusion presented in \cite{Gong2020} where an asymmetry in the pulsation of the 1227 s signal is visible. However, our data is supplemented by the 2021 observation which is on-axis and captures the source with the PN detector.

\begin{deluxetable*}{lcccccccc}[ht]
\tablecaption{Summary of \xmm\ EPIC and NuSTAR Timing Analysis}
\tablecolumns{6}
\tablehead{
\colhead{Instrument}
&
\colhead{{Peak frequency [mHz]}}  
&
\colhead{Peak period [sec]}
&
\colhead{Counts}
&
\colhead{ $Z^2_3$ }
& 
\colhead{{Significance ($\sigma$)$^b$}} 
&
\colhead{$P $ [\%]$^a$}  
}
\startdata   
\xmm\ 0.3--5 keV & & &  & &  & \\
 \hline
\textbf{0553950201} \\
MOS & $1.6288_{-0.0002}^{+0.0006}$ & $613.94_{-0.22}^{+0.08}$ & 2636 & 115.77  & $>$ 7 & 30 $\pm$ 2 \\
\textbf{0870990201}\\
MOS & $1.6260_{-0.0030}^{+0.0016}$ & $615.01_{-0.61}^{+1.13}$  & 1156 & 78.56 & 7.0 & 37 $\pm$ 3 \\
PN  & $1.6277_{-0.0026}^{+0.0025}$ & $614.37_{-0.95}^{+0.97}$ & 1114 & 75.74  &  6.9 & 37 $\pm $ 3 \\
\hline
\hline
\nustar\ 3--5 keV &&&&& \\
\hline
\textbf{30601021002}\\
FPMA + FPMB & $1.6317_{-0.0025}^{+0.0010}$ & $612.84_{-0.36}^{+0.58}$ & 768 &  23.09 & 1.4 & $<32$ \\
\hline
\enddata
\label{tab:timing}

 All errors shown are at 90\% confidence intervals. The results are from the $Z^2$ test in the narrow frequency band. For the 3--5 keV \nustar\ data from which the 614-second signal was not significantly detected, the peak frequencies/periods (without error bars) and the  upper limits on the pulsed fraction are listed. The total counts for each observation, instrument and energy band are listed as well.  

$^a$Pulsed fractions [\%] and upper limits were calculated by \textbf{HENzsearch}  \citep{matteo_bachetti_2022_6394742}. All of the hard-energy band $Z^2_3$ searches produced no significant peaks above 1-$\sigma$. 

\end{deluxetable*}

For \nustar\ data, we applied the barycentric correction to the extracted  photon events using the \textbf{barycorr} command line tool from HEASoft Version 6.25 \citep{heasoft}. We searched for a periodic signal in \nustar\ data using the same $Z^2_3$ test and a step-size of $\Delta f = 2\times 10^{-6}$ between $f=1-2$ mHz where we expect the 614-sec signal and its higher harmonics.  A broadband 3--10 or 3--40 keV search in the frequency band of  $f=10^{-4} - 2$ Hz yielded no significant detection except a peak at $P\sim32.3$ min (with $5-\sigma$ significance). However, the 32.3 min peak is likely a harmonic of the \nustar's 96.8 min orbital period as evidenced by the lack of a counterpart in the \xmm\ data.
In Table 2, we present our 3--5 keV \nustar\ results since the 614-sec modulation was detected by \xmm\ below 5 keV. We found no significant pulsations and the $P\sim614$ sec signal appeared to be at $1.4\sigma$ level.  
The insignificant detection of the 614-sec periodicity, with a pulsed fraction $<32$\%, is likely due to the poor statistics  at $E<5$ keV where \xmm\  observations demonstrated that the signal was most pronounced.    
    
\section{X-ray spectral analysis} 

The \nustar\ spectra of the source (ObsID 30601021002) held $\sim$ 2,200 combined FPMA + FPMB counts in the 3--50 keV band. We jointly fit the \nustar\ spectra with the 2021 \xmm\ observation data (with $\sim$ 3,000 combined MOS1, MOS2 and PN counts in the 0.5--10 keV band) and the 2010 \xmm\ observation data (with $\sim$ 2,800 combined MOS1 and MOS2 counts in the 1.5--10 keV band). The MOS spectra in the 2010 \xmm\ observation is dominated by background below $\sim1.5$ keV. 

We first fitted a handful of phenomenological models to characterize the overall X-ray spectral properties using {\tt XSPEC} \cite{Arnaud1996}. Each spectral model was multiplied by {\tt tbabs} (for the ISM absorption) using the Wilms abundance data \citep{Wilms2000}. Furthermore, we used  a cross-normalization factor between the \xmm\ and \nustar\ spectra using {\tt constant} in {\tt XSPEC}. Figures 3 and 4 present the best model fits and residuals for (1) a  power-law model; (2) a power-law fit with a Gaussian line  component at $\sim6-7$ keV to account for Fe emission lines; (3) an {\tt APEC} model fit; (4) an {\tt APEC} model with a Gaussian line component fixed at $E=6.4$ keV and (5) an {\tt APEC} model with a Gaussian line at 6.4 keV and a {\tt pcfabs} partial absorption component. A single power-law model yields a poor fit to the \xmm\ and \nustar\ spectra with a  $\chi^2_\nu = 1.41$ (380 dof) largely due to the residuals at $E\sim6-7$ keV. Adding a Gaussian line model  improved the fit to $\chi^2_\nu = 1.15$ (377 dof). The best-fit line energy of 6.65 keV and large equivalent width of 940 eV indicates a complex Fe K emission line mostly originating from H- and He-like ions.  The X-ray source spectra are apparently characterized by thermal emission. Similarly, an absorbed  blackbody model  yielded  a poor fit with $\chi_\nu^2 \sim 2$. 

We then fit an absorbed {\tt APEC} model, i.e. a single-temperature optically-thin thermal plasma model \citep{Smith2001}. The best-fit plasma temperature and abundance are $kT = 33_{-7}^{+12}$ keV and $Z = 2.2_{-0.8}^{+1.5} \:Z_\odot$, respectively. The fit is further improved by adding a narrow Gaussian line fixed at $E=6.4$ keV indicating the presence of a neutral Fe K-$\alpha$ line. Furthermore, we added a partial covering absorption model ({\tt pcfabs}), improving the fit of low-energy bins. Although unaccounted for in \cite{Gong2022}, {\tt pcfabs} is needed to model the X-ray absorption due to the accretion curtain that IPs exhibit (\cite{Norton1989}, \cite{Hailey2016}). The {\tt pcfabs*(APEC+gauss)} model, while freezing the gaussian line at $E=6.4$ keV, represents the best-fit case with $\chi_\nu^2 = 1.00$ (376 dof). The plasma temperature is still high at $kT = 25\pm6$ keV. Our fit results are summarized in Table 3. Through the best-fit flux normalization factors, we found that the X-ray flux decreased by $\sim 15$\% from 2010 to the 2021 observations. The neutral hydrogen column density is $N_{\rm H} \sim4\times10^{22}$ cm$^{-2}$ for all the models except when we applied the partial covering absorption model (which yielded a lower value of $N_{\rm H} = 1.6 \times10^{22}$ cm$^{-2}$). 

\begin{deluxetable*}{lccccc}[ht]
\tablecaption{Phenomenological model fits to \xmm\ EPIC and \nustar\ spectra}
\tablecolumns{6}
\tablehead{
\colhead{Parameter}   
&
\colhead{{\tt pow}}  
& 
\colhead{{\tt pow + \tt gauss}}
& 
\colhead{{\tt APEC}}
&
\colhead{{\tt APEC}+{\tt gauss}}
&
\colhead{{\tt pcfabs}*({\tt APEC}+{\tt gauss})}
}
\startdata  
$C_{2021}^a$ & $0.87_{-0.06}^{+0.07}$ & $0.87_{-0.06}^{+0.07}$ & $0.84 \pm 0.06$ & $0.85\pm 0.06$ & $0.87 \pm 0.06$ \\
$C_{2010}^a$ & $1.11_{-0.09}^{+0.10}$ &  $1.14 \pm 0.09$ & $1.09_{-0.08}^{+0.09}$ & $1.11_{-0.08}^{+0.09}$ & $1.18 \pm 0.09$ \\
$N^{(i)}_H (10^{22} \rm{cm}^{-2})^b$ & $4.6\pm0.4$ & $4.1\pm0.4$ & $4.2\pm0.3$ & $4.0\pm0.3$ & $1.6\pm0.5$\\
$N^{(p)}_H (10^{22} \rm{cm}^{-2})^c$ & ... & ... & ... & ... &   $7.7_{-1.6}^{+2.4}$ \\
$c^c_f$ & ... & ... & ... & ... & $0.77_{-0.09}^{+0.08}$ \\ 
$\Gamma$ & $1.48 \pm 0.08$ & $1.46 \pm 0.08$ & ... & ... & ... \\
$kT$ (keV) & ... & ... & $33.0_{-6.5}^{+11.6}$ & $37.3_{-9.1}^{+15.3}$ & $24.7_{-5.6}^{+5.8}$ \\
$Z^d (Z_\odot)$ & ... & ... & $2.2_{-0.8}^{+1.5}$ & $2.6_{-1.1}^{+1.7}$ & $1.1_{-0.5}^{+0.7}$ \\
$E_{line}$ (keV) & ... & $6.65_{-0.10}^{+0.09}$  & ... & $6.4^*$ & $6.4^*$ \\
$\sigma_{line}$ (keV) & ... & $0.43_{-0.13}^{+0.26}$ & ... & $0.01^*$ & $0.01^*$ \\
$EW_{line}$ (eV) & ... & $937_{-114}^{+66}$ & ... & $188_{-10}^{+59}$ & $146_{-37}^{+35}$ \\
$F_X (10^{-12} $\eflux$)^e$ & $3.91 \pm 0.02$ & $3.88 \pm 0.02$ & $3.34_{-0.04}^{+0.03}$ & $3.41 \pm 0.04$ & $3.33 \pm 0.05$ \\ 
$\chi^2_{\nu}$ (dof) & 1.41 (380) & 1.15 (377) & 1.26 (379) & 1.18 (378) & 1.00 (376) \\
\enddata
\label{tab:prelimfits}
All errors shown are 1-sigma intervals. \\
$^a$ Cross-normalization factors of the \xmm\ data in 2021 ($C_{2021}$) and in 2010 ($C_{2010}$) with respect to the \nustar\ observation in 2021. \\
$^b$ The ISM hydrogen column density associated with {\tt tbabs} which is multiplied to all the models. \\
$^c$ $N^{(p)}_H$ and $c_f$ are the hydrogen column density and covering fraction associated with {\tt pcfabs}. \\
$^d$ Abundance relative to solar. \\
$^e$ Unabsorbed 3--50 keV flux of the \nustar\ data. \\
$^*$ The parameter is frozen. 

\end{deluxetable*}

\begin{figure}[ht]
\begin{minipage}{.6\textwidth}

 \includegraphics[width=10cm]{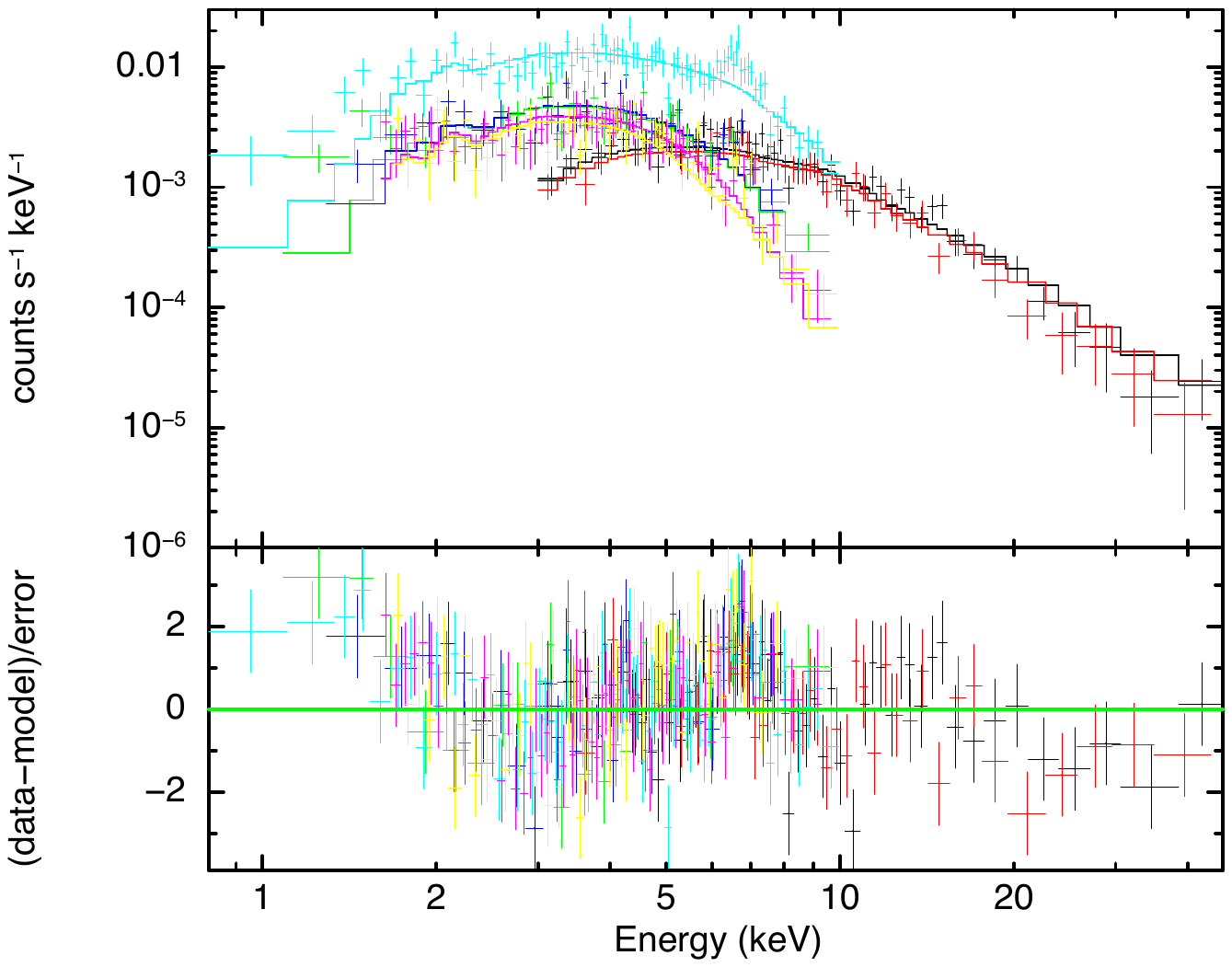}
 
\end{minipage}%
\begin{minipage}{.6\textwidth}
  \hspace{-2cm}
  \includegraphics[width=10cm]{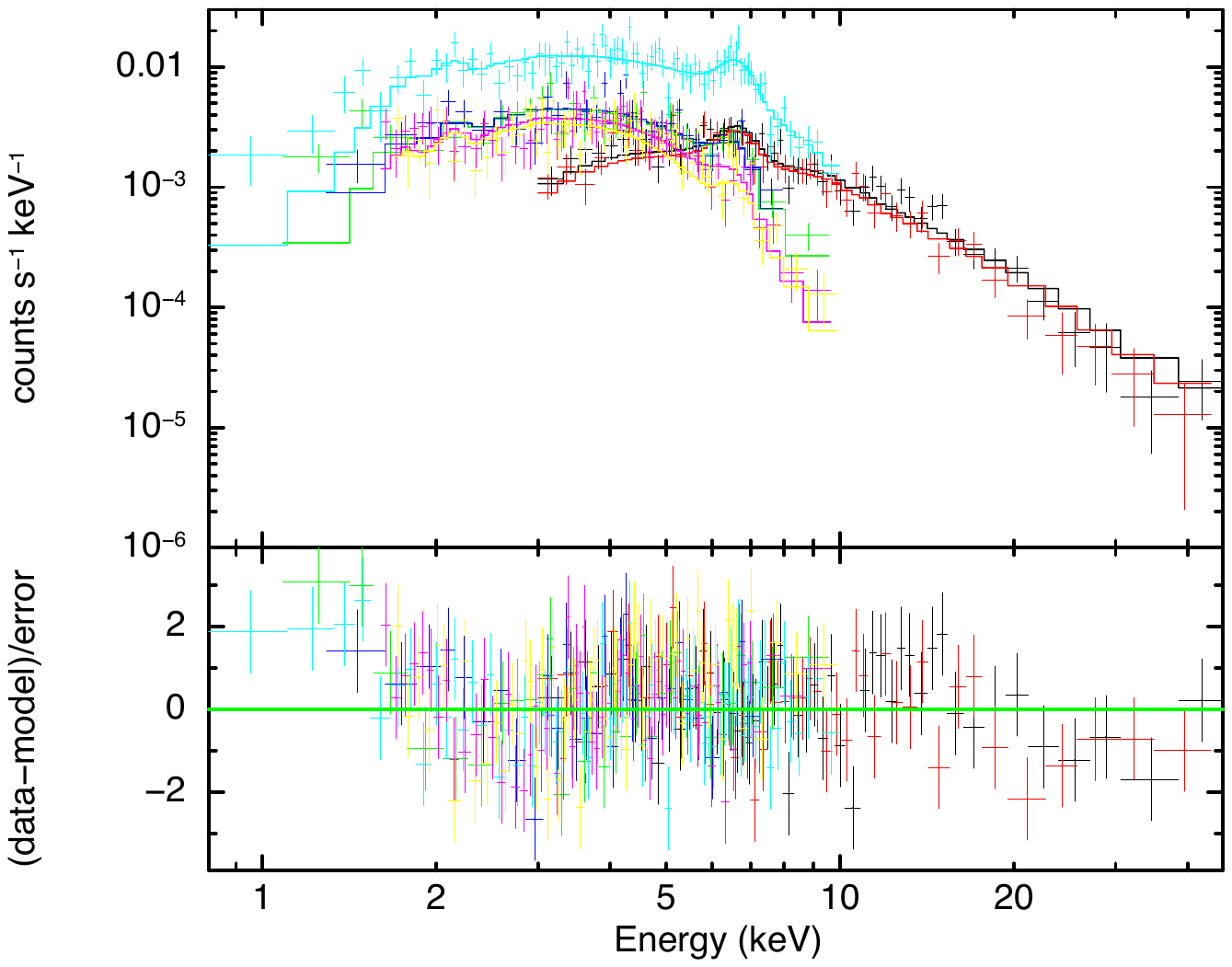}
\end{minipage}
\begin{minipage}{.6\textwidth}
 \includegraphics[width=10cm]{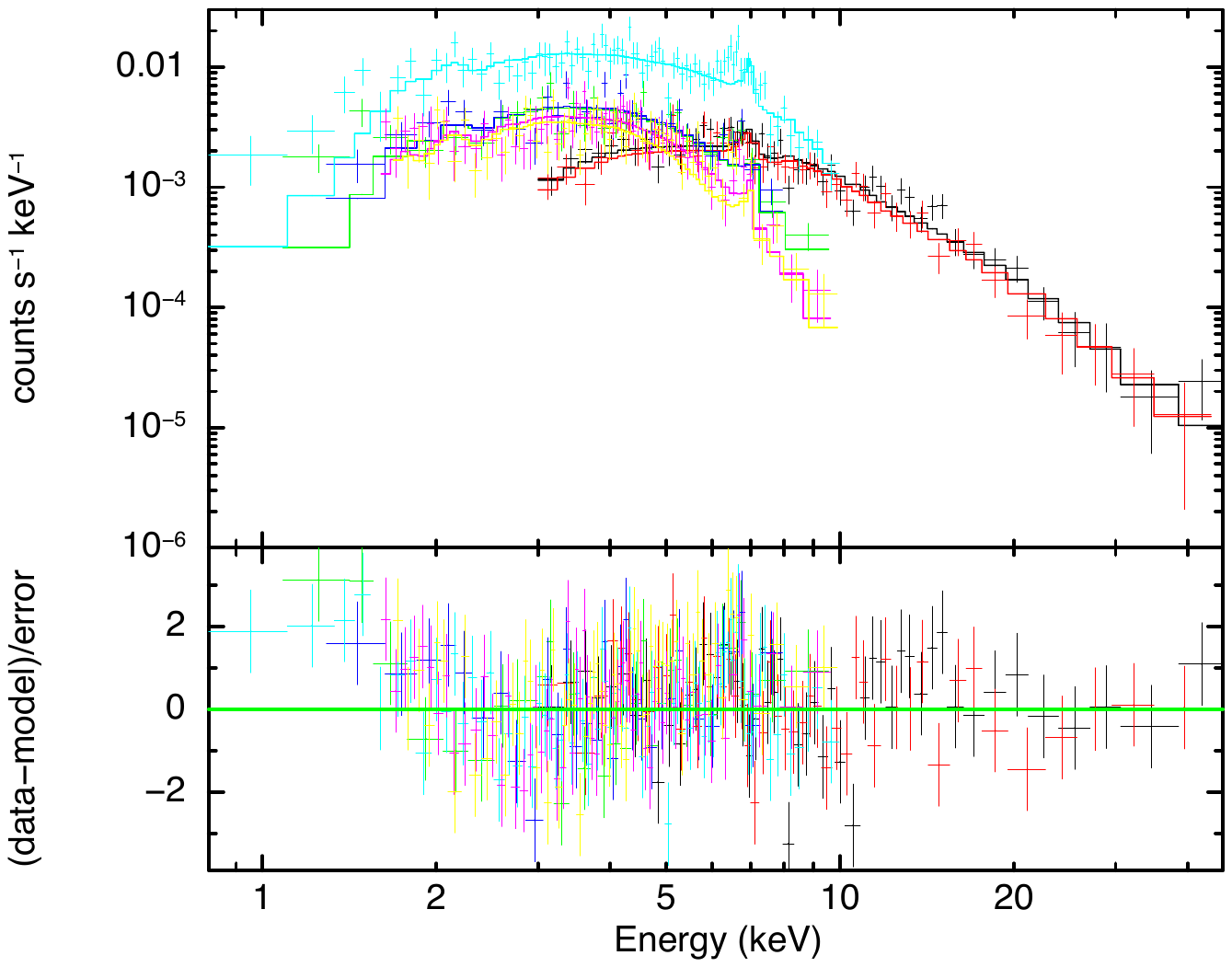}
\end{minipage}%
\begin{minipage}{.6\textwidth}
  \hspace{-2cm}
  \includegraphics[width=10cm]{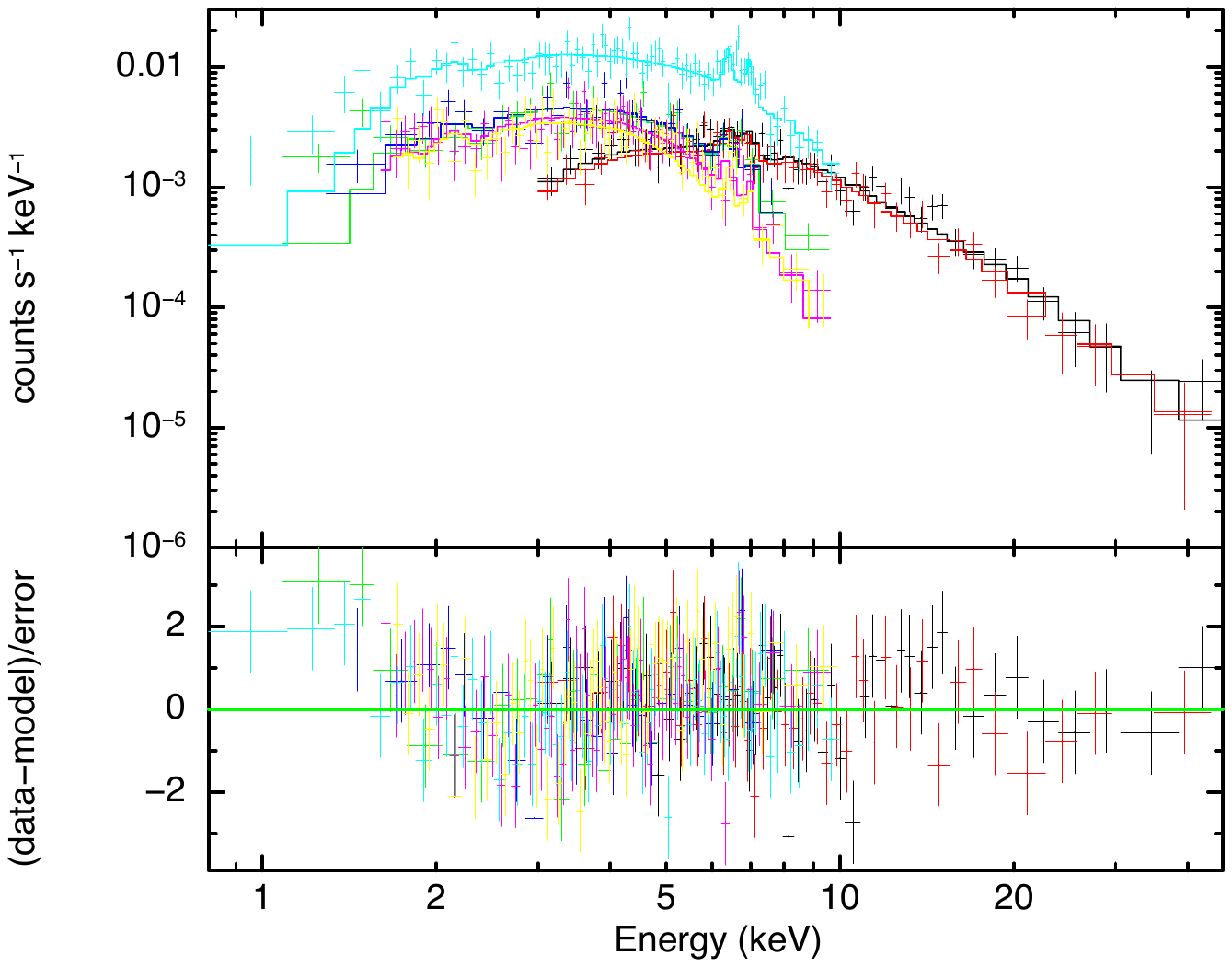}
\end{minipage}
  \caption{\xmm\ EPIC and \nustar\ spectra fit with four phenomenological models. From top left to top right: (1) an absorbed power-law model fit; (2) an absorbed power-law fitted with a Gaussian at 6.4 keV component; (3) an absorbed {\tt APEC} model fit; (4) an absorbed {\tt APEC} model with a Gaussian at 6.4 keV component. The $\nustar$ observation used here is 30901021002 and is presented in the 3 -- 50 keV band, with FPMA (1,146 counts) in black and FPMB (1,097 counts) in red.  The $\xmm$ observation 0553950201 is presented in the 1.5 -- 10 keV band with MOS1 (1,438 counts) in magenta and MOS2 (1,342 counts) in yellow. The more recent $\xmm$ observation 0870990201 is presented in the 0.5 -- 10 keV band with MOS1 (667 counts) in green, MOS2 (671 counts) in navy blue and PN (1,688 counts) in light blue. }
\label{fig:spectra_phenomeno}
\end{figure}

\section{Optical and infrared Counterparts}

We found no optical counterpart within $\sim 15$\asec\  of \src\ \citep{GaiaED32020} due to high ISM absorption. 
\cite{Greiss2014} identified a possible IR counterpart to \src\ by cross-referencing the \chandra\ Galactic Bulge Survey and the VISTA Variables in the Via Lactea (VVV) Survey \citep{Greiss2014}. However, the IR counterpart (VVV J174517.24-321357.67) is offset by $3$\asec\ from \src\, and, consequently, the match between the two sources has a large false-alarm probability of 0.6 \citep{Greiss2014}. 
We, therefore, consider no IR counterpart was detected for \src. The source is thus likely fainter than the sensitivity limits for the VVV Survey. We used the sensitivity limits of the VVV Survey to constrain the spectral type of the companion star. The magnitude limits for the VVV Survey are $J=20.2,\, H=18.2,$ and $K_{s}=18.1$. Using the relative extinctions from the VVV Survey \citep{Saito2012} and assuming $E(B-V)=1.8$ (the typical extinction value in the direction of the Galactic Bulge) \citep{Greiss2014}, we found the $J$, $H$, and $K_s$ extinctions (1.56, 1.02, and 0.655, respectively) with an optical extinction $A_V=5.55$. Assuming a source distance of 8 kpc \citep{Wevers2016}, we derived absolute magnitudes of $M_J=4.125,\, M_H=2.665,\, \text{and }M_{K_s}=2.930$. Using the $M_{K_s}$ limit value, we find a main sequence F9/F9.5 spectral type or later \citep{Pecaut2013}.  Using X-ray observations, \citet{Jonker2014} establishes a lower bound of 3 kpc for the source distance. If the true source distance were at this lower bound, the extinction would be lower than $E(B-V)=1.8$. A lower extinction would result in an even fainter spectral type than the above results. Therefore, regardless of distance, we conclude that the companion star of \src\ is a main sequence star of type F9/F9.5 or later.  

\section{Discussion} \label{sec:discussion}

In this section, we first discuss the source type of \src\ primarily based on the X-ray spectral properties (\S6.1). Then, after identifying \src\ as an IP, we measure its WD mass using our X-ray spectral model for mCVs  (\S6.2) and discuss some implications of characterizing the mass and B-field of an IP at distances (d $\simgt$ 3 kpc) where optical observations are hampered by significant extinction(\S6.3).  

\subsection{Source identification}

We present our source identification scheme for \src\ based on the X-ray analysis results and counterpart searches in other wavelengths. The lack of a bright IR counterpart (\S 5) rules out the possibility that \src\ is a high-mass X-ray binary. The strong Fe K emission lines suggest that the source is not a low-mass X-ray binary whose X-ray emission is non-thermal (occasionally with weak Fe lines) \citep{Mori2021}. The presence of Fe emission lines as well as the long-term variability rules out a pulsar origin. Below we consider two possible source types suggested by \citet{Gong2022} and conclude that the X-ray source is most likely an IP mainly based on the X-ray spectral properties obtained by the \xmm\ and \nustar\ observations. 
\subsubsection{UCXB} 

Solely based on the original 2010 \xmm\ observation, it was proposed that \src\ may be a UCXB with a short orbital period of 614 sec \citep{Gong2022}. UCXBs are low-mass X-ray binaries usually containing a WD donor and a neutron star accretor (or very rarely a stellar-mass black hole) and with an ultra-short ($< 1$ hour) orbital period. X-ray spectra of known NS-UCXBs are well characterized by a $kT \sim$ 1--3 keV blackbody component originating from the NS boundary layers \citep{Koliopanos2015}. 
We found a UCXB origin implausible  because an absorbed blackbody model did not fit the 0.5 -- 50 keV \xmm\ + \nustar\ spectra well. Furthermore, other combinations with a blackbody model (e.g., blackbody + power-law) did not fit the X-ray spectra due to the prominent Fe lines at $E = 6-7$ keV. 
In addition, the known UCXBs exhibit no or weak Fe-K emission lines in their X-ray spectra \citep{Koliopanos2020}. UCXBs with He-rich donors show weak Fe lines with equivalent width (EW) $\approx$ 50--80 eV (except for X-ray transients, which show stronger Fe-K features), whereas those with C/O or O/Ne/Mg-rich donors show no traces of Fe emission lines  \citep{Koliopanos2020}. Note that these weak or no Fe line features were observed from a dozen very bright or outbursting UCXBs with $L_{\rm X} > 10^{36}$ erg\,s$^{-1}$.  For a much fainter UCXB similar to \src\ ($L_{\rm} \simlt 10^{34}$ erg\,s$^{-1}$), Fe line emission should be negligible due to a lack of bright illuminating X-rays. 
Thus, the strong Fe emission lines detected with a total EW $\sim$ 0.9 keV also rule out the UCXB hypothesis. 

\begin{figure}[ht]
\centering
\begin{minipage}{.9\textwidth}
  \centering
 \includegraphics[width=12cm]{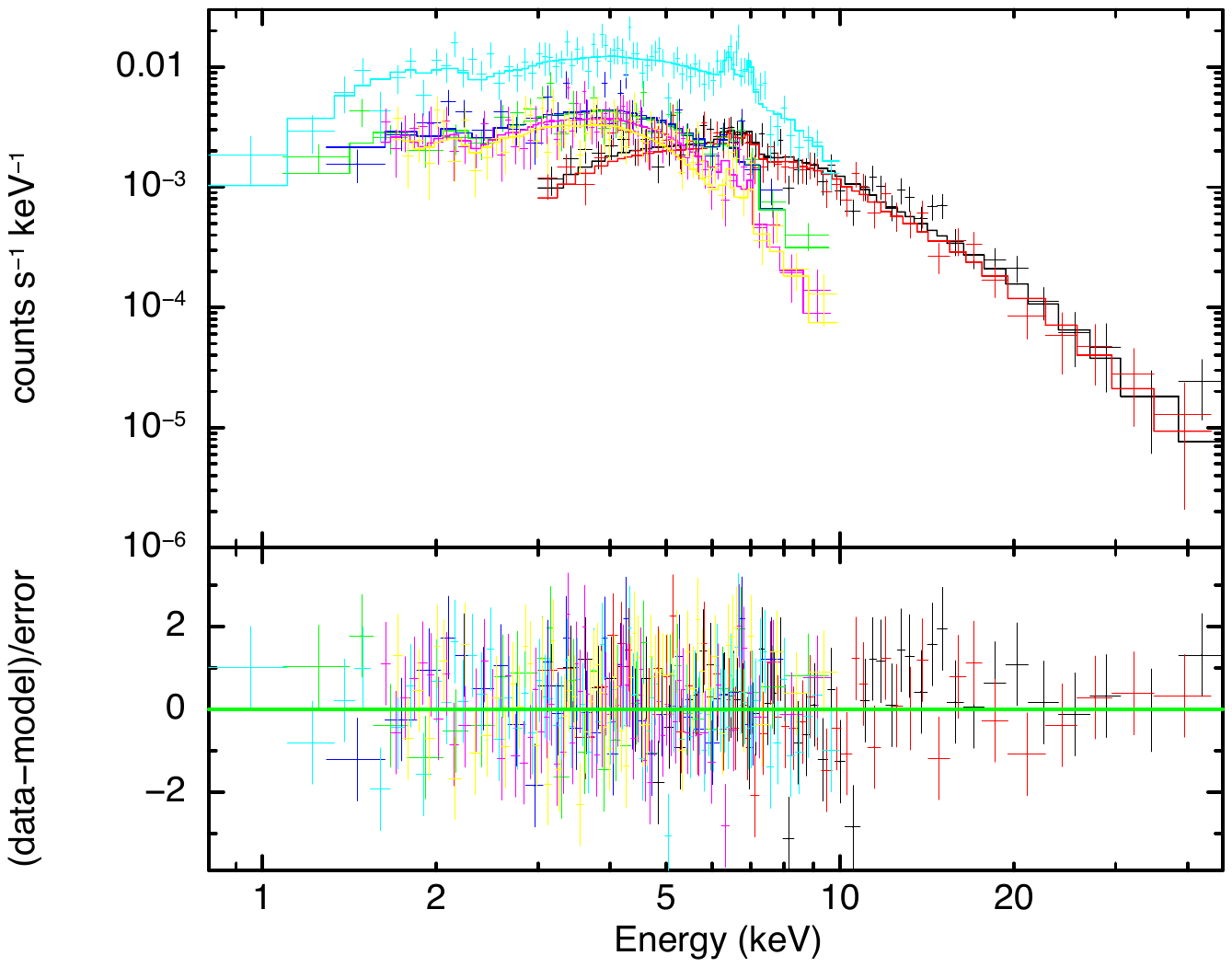}
\end{minipage}
\caption{\xmm\ and \nustar\ spectra and residuals of \src\ fit with an  {\tt tbabs}*{\tt pcfabs}*({\tt APEC}+{\tt gauss}) model, which yielded a $\chi^2_{\nu} = 1.00$ for 376 degrees of freedom, and which strongly supports the IP scenario. The Gaussian line fitted at 6.4 keV and modeling the Fe K emission line yielded an equivalent width of $146_{-49}^{+68}$ eV, which is typical of IPs.} 
\end{figure}

\subsubsection{Intermediate polar} 

The variable, thermal X-ray spectra with atomic emission lines indicate  that \src\ is a CV. In particular, the hard X-ray spectra extending up to $E\sim50$ keV and hot plasma temperature $\sim 30-40$ keV suggests an IP origin.  
IPs are copious emitters of hard X-rays above 20 keV as a large fraction of the  hard X-ray sources detected by $\INTEGRAL$ have been identified as IPs  \citep{Masetti2010}. The typical IP plasma temperature is $kT \sim 20-40$ keV \citep{Mukai2017}. On the other hand, the other types of CVs, namely polars and non-magnetic CVs (nmCVs or dwarf novae), exhibit much softer and fainter X-ray spectra. Polars, with typically higher magnetic fields ($B \sim 10 - 200$ MG) than those of IPs, have lower plasma temperatures ($kT \sim 5 - 20$ keV) due to faster cyclotron cooling \citep{Mukai2017, Warner2003}. On the other hand, nmCVs are characterized by even fainter and softer X-ray emission with $kT \simlt 10$ keV \citep{Byckling2010}. Even a few outliers with harder X-ray spectra (e.g., SS Cyg with $kT \sim$ 20 keV during quiescence \citep{Ponman1995})  do not reach the plasma temperature as high as those of \src. 
Furthermore, nmCVs tend to be far less luminous than IPs, with typical 0.25 -- 10 keV luminosity ranging anywhere from $10^{30}$ erg s$^{-1}$ to $10^{32}$ erg s$^{-1}$ \citep{Wada2011}. The measured, unabsorbed X-ray flux of $\sim1.4\times10^{-12}$ \eflux\ converts to $L_{\rm X} = 2\times10^{33} - 1\times 10^{34}$ erg\,s$^{-1}$ assuming the source distance of 3--8 kpc. This X-ray luminosity is much higher than those of nmCVs.    
Moreover, the measured EW for the Fe K-$\alpha$ line at 6.4 keV is 188 eV and 146 eV for {\tt APEC+gauss} and {\tt pcfabs*(APEC+gauss)} models, respectively (Table 2). These results are consistent with the observed mean EW of neutral Fe K-$\alpha$ line of 115 eV for IPs and are markedly different from the polars' and nmCVs' of 68 and 62 eV, respectively \citep{Xu2016}. The best-fit model (Figure \ref{fig:spectra_phenomeno}) requires a partial-covering absorption term, and this is also consistent with the IP hypothesis because X-ray emission from the accretion column is partially shielded by the  accretion curtain or reflected from the WD surface (\cite{Norton1989}, \cite{Hailey2016}). In addition, the 614 sec period is in the typical spin period range of IPs (see Table 3 in \citet{Suleimanov2019}), and the low energy X-ray spin modulation is often observed among IPs due to the accretion curtain absorption \citep{Anzolin_2008}. While double-peaked modulation is a common feature of IPs, the non-overlapping, highly symmetric pulses in the $\sim $1227 sec and $\sim $1841 sec regions indicate that the $\sim$ 614 sec signal is a single-peaked modulation. While this is typically associated with slower-spinning IPs, this feature has been previously detected in faster-spinning IPs ($\mathrm{P}_{\rm spin}$ = 564 sec, 450 sec) \citep{Bonnet_Bidaud_2007, Gorgone_2021}. This indicates that the emissions of only one of the two WD poles are visible in the observations. Lastly, radio observations of the local position of J1745 with ATCA in Australia in late 2021 provided inconclusive detection of the source, while later VLA observations yielded no detection in the radio band at all \citep{Gong2022}. The lack of radio emission is also congruent with the IP hypothesis since \citet{Barrett2020} reported that only a small fraction of IPs had been detected in the radio band.
 
\subsection{Determining the White Dwarf Mass of J1745}

 In this section, we outline our methodology for determining the WD mass of \src. In general, X-ray emission from IPs originates primarily from the accretion column. Some X-rays are reprocessed via absorption in the accretion curtain and reflection by the WD surface (\cite{Norton1989}, \cite{Hailey2016}). We utilize the broad-band X-ray spectra to constrain the parameters such as the magnetospheric radius, accretion rates, and WD mass.
 
\begin{figure}[ht]
\begin{center}
\includegraphics[width=8.7cm,angle=0]{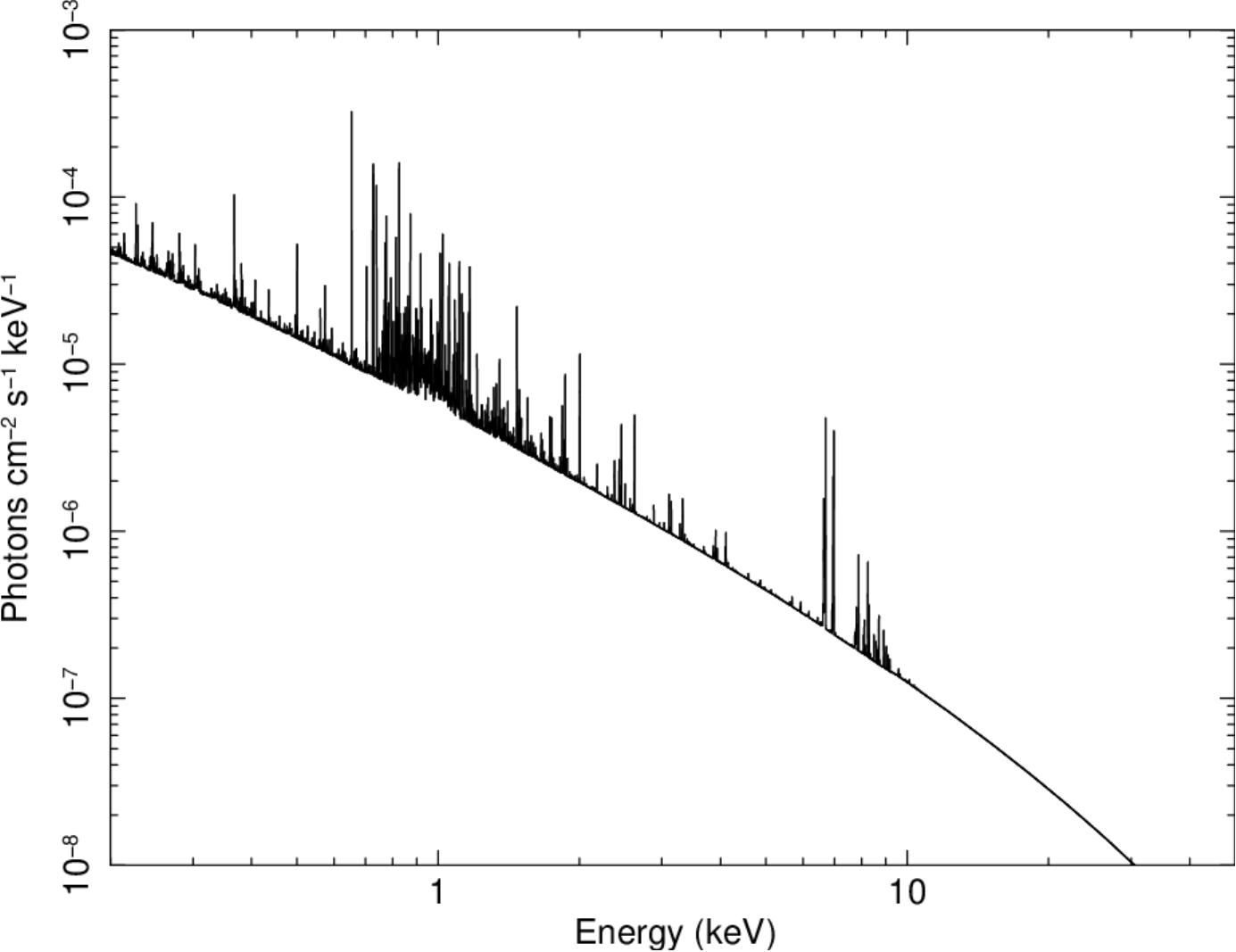}
\end{center}
\caption{{\tt MCVSPEC} model spectrum with $M = 1 M_{\odot}$ and $\dot{m} = 1$ g\,cm$^{-2}$\,s$^{-1}$ and $R_m/R = 20$.}
\label{fig:mcvspec_model} 
\end{figure}

 To model X-ray emission from the accretion column, we developed an XSPEC model based on the accretion flow equations presented in \citet{Saxton2005}. The {\tt XSPEC} model, called {\tt MCVSPEC}, assumes a steady, planar accretion flow channeled along B-field lines from the shock height to the WD surface (Mori et al. in preparation). In IPs, the accretion flow begins free-falling along the magnetic field lines from the magnetospheric radius ($R_m$) defined by the balance between magnetic and gas pressures. For a given set of $M$, $R_m$, and specific accretion rate \mdot\ [g\,cm$^{-2}$\,s$^{-1}$] taken as input parameters, coupled differential equations for the flow continuity and momentum/energy conservation are numerically solved along the accretion column to determine the density and temperature profiles. The model considers both thermal bremsstrahlung and cyclotron cooling, the latter of which is prevalent in polars or highly magnetized IPs with $B\simgt 30$ MG \citep{Wu1994}. An X-ray model spectrum is calculated by integrating the emissivity (with different $kT$ and $\rho$ values) throughout the accretion column by implementing the most updated atomic database for collisionally ionized plasma (\url{ http://atomdb.org}). A significant improvement in our modeling relevant to IPs, and missing from the analysis of \cite{Gong2022}, is the incorporation of the recent observation \citep{Suleimanov2016} that the accreting gas begins its infall from the disk magnetospheric radius, not from infinity, as was previously assumed in all models. As pointed out by \cite{Suleimanov2016}, if the finite magnetospheric radius effects for IPs are not taken into account, it will lead to an overestimation of the free-fall velocity at the shock height since $v_{ff} = \sqrt{2GM (\frac{1}{R+h} - \frac{1}{R_m})}$. Furthermore, since $v_{ff}$ is directly linked to the shock temperature, WD masses in IPs will be underestimated if the effect of the finite magnetospheric radius is ignored.  Furthermore, {\tt MCVSPEC} accounts for the effects of finite shock height by a self-correcting shooting method that adjusts the free-fall velocity and other appropriate parameters to the model's derived shock height until it converges.
For example, Figure \ref{fig:mcvspec_model} shows an {\tt MCVSPEC} model spectrum with WD mass $M = 1 M_{\odot}$ and $\dot{m} = 1$ g\,cm$^{-2}$\,s$^{-1}$ and $R_m/R = 20$. 

In the following subsections, we outline how we constrain $R_m$ (\S 6.2.1) and $\dot{m}$ (\S6.2.2 and 6.2.3) prior to the spectral fitting and eventually measure the WD mass (\S 6.2.4). 
The process begins by deriving $R_m/R$ from the spin period, assuming that the IP is in a spin equilibrium. {\color{black} We then bound the specific accretion rate by considering a range of the source distance ($d$) and fractional accretion column area ($f$). Following these parameter constraints, we proceed to X-ray spectral fitting with the {\tt MCVSPEC} and X-ray reflection model, where some of the input parameters (e.g., a reflection fraction) are properly linked with each other or frozen to the pre-determined values (e.g., $R_m/R$). As described below, our physically motivated model allows us to constrain the IP parameters. To clarify our method, we provided a flowchart in the appendix.} 

\subsubsection{Magnetospheric radius} \citet{Suleimanov2019} suggested that the frequency of a spectral break in the power density spectrum could be used to estimate $R_m$. However, besides the spin period of 614 sec and its harmonics, we found no other feature or break in the power density spectra obtained from the \xmm\ and \nustar\ data. Therefore, following \citet{Shaw2020}, we estimated $R_m$ assuming that the WD is in  spin equilibrium with the accretion disk. We derived $R_m/R = 20\rm{-}22$ for $M = (1\rm{-}1.4) M_\odot$ 
using $R_m = \left(\frac{GMP^2}{4\pi^2}\right)^{1/3}$ where $P$ is the WD spin period (614 sec). Note that the dependence on $M$ is small compared to $P$, so we adopt the mass estimate of \cite{Gong2022}. The ratio of $R_m/R\sim20$ is within the range of $R_m/R = 8\rm{-}25$ obtained by a systematic study of 45 IPs using \nustar\ and \swift-BAT data \citep{Suleimanov2019}. Hereafter we adopt $R_m/R = 20$ for {\tt MCVSPEC}. 

\subsubsection{Mass accretion rate and source distance} 

{\color{black} The total mass accretion rate $\dot{M}$ [g\, s$^{-1}$] can be estimated from the bolometric luminosity ($L$): 
$L =  GM \dot{M} \left(\frac{1}{R} - \frac{1}{R_m}\right)$. Since the source was not detected below $\sim1$ keV, we set $L\approx L_X$, for now, where $L_X$ is the absorption-corrected  X-ray luminosity, assuming that most of the radiation is emitted in the X-ray band through thermal bremsstrahlung emission. We used $L_X = 4 \pi d^2 F_X$ where $F_X$ is the unabsorbed X-ray flux in 0.8--50 keV and assumed source distance ($d$).  
However, in reality, $L > L_X$ because some X-rays emitted from the accretion column can be reprocessed in the lower energy band ($E\simlt1$ keV). For example, a soft X-ray blackbody component of $kT = 86$ eV constitutes $\sim30$\% of the whole X-ray emission for IP 1RXSJ154814.5-452845 \citep{Haberl2002}. Furthermore, when the WD B-field is high, magnetic CVs can have a non-negligible fraction of the radiation through cyclotron emission in the optical and UV band. Since no optical or UV emission from the source at a large distance ($d > 3$ kpc) is observable, we need to rely on modeling the ratio of thermal bremsstrahlung and cyclotron cooling rates ($\epsilon = L_{\rm cyc}/L_{\rm br}$). $\epsilon$ is a function of the accretion column height ($h$) varying with the plasma density and temperature. In general, $\epsilon(h)$ is the highest at the shock height ($\epsilon_s$) where the plasma density and temperature are lowest and highest, respectively \citep{Wu1994}. Based on our spectral fits and the range of the specific accretion rate, as described below, we found that $\epsilon_s \sim 0.1\rm{-}0.5$ and that $\epsilon$ rapidly decreases toward the WD surface. To estimate the overall contribution of the cyclotron cooling component in the bolometric luminosity, we calculated the mean ratio ($\bar\epsilon$) by integrating $\epsilon$ over the accretion height  (from $h=0$ to $h_s$), using the best-fit temperature and density profile obtained by {\tt MCVSPEC}, and dividing by $h_s$.  
For each case of the model fitting, $\bar\epsilon$ allows us to estimate how much luminosity we are missing by considering the X-ray flux measurements. 
If the cyclotron emission is not negligible, the accretion rate is underestimated by assuming $L \approx L_X$, especially for higher $M$ and $\dot{M}$ values which yield greater $\bar{\epsilon}$ (see \citet{Wu1994} for how $\epsilon_s$ depends on $M$ and $\dot{M}$). In the subsequent sections, we find that $\bar{\epsilon}$ is at most $\sim20$\% and describe how the cyclotron emission component affects our IP parameter determination. 
Hereafter, we adopt a range of $\dot{M} = (2\rm{-}10) \times 10^{16}$ [g\,s$^{-1}$], corresponding to $d = 3\rm{-}8$ kpc, respectively, assuming $L \approx L_X$. We will later discuss how the underestimation of the mass accretion rate due to the ``missing" luminosity in the soft X-ray band and below will affect the WD mass measurements.}
 
\subsubsection{Specific mass accretion rate}

Given that $R_m/R$ and $\dot{M}$ are constrained, we proceed to bound the specific accretion rate ($\dot{m}$) by considering the fractional accretion column area ($f$) in two extreme cases. 
Note that the specific accretion rate is defined as $\dot{m} = \frac{\dot{M}}{4 \pi R^2 f}$, where $f$ is the fractional area of the accretion column base with respect to the WD surface. In general, $f$ is unknown observationally except for an eclipsing IP XY Ari with $f < 2\times10^{-3}$  \citep{Hellier1997}.  
We first set an upper limit of $f$ to $f_{max} = \frac{R}{2R_m} = 2.5\times10^{-2}$. This is the maximum fractional area of the accretion column for a dipole B-field geometry assuming that the infalling gas originates from the entire accretion disk \citep{2002apa..book.....F}. 
{\color{black}A combination of $d = 3$ kpc and $f_{max}$ leads to a minimum specific accretion rate of $\dot{m} = 0.6$ g\,cm$^{-2}$\,s$^{-1}$. This value sets a lower limit of $\dot{m}$ since, in reality, the mass accretion  originates from the so-called threading region where the gas in the inner accretion disk is connected with the magnetic field lines toward the polar regions \citep{Belloni2021}, in contrast to emission coming from the entire disk (corresponding to $f=f_{max}$). For a lower limit of $f$, we adopted the theoretical range of $0.001 \simlt f \simlt 0.02$ from \citet{Rosen1992} who considered various IP parameters and geometrical configurations. Using the theoretical lower bound of $f = 0.001$ and $d = 8$ kpc, we derive the maximum specific accretion rate of $\dot{m} = 44 $ g\,cm$^{-2}$\,s$^{-1}$. Hereafter, we present our fit results for the range of $\dot{m} = 0.6\rm{-}44$ g\,cm$^{-2}$\,s$^{-1}$.}
 
\subsubsection{Spectral fit results}  

Given that two of the input model parameters ($R_m/R$ and $\dot{m}$) are constrained as above, we are able to fit {\tt MCVSPEC} to the \xmm\ and \nustar\ spectra with less parameter degeneracy. In contrast to \cite{Gong2022}, we also  considered additional photon reprocessing effects associated with X-ray emission from IPs. Some X-rays emitted from the accretion column can be absorbed in the pre-shock region or reflected by the WD surface as manifested in the neutral Fe fluorescence lines \citep{Hayashi2021}.   
Following previous X-ray studies \citep{Hailey2016}, we add a gaussian line component at $E=6.4$ keV for the neutral Fe K-$\alpha$ fluorescence line, {\tt pcfabs} in {\tt XSPEC} to account for X-ray absorption at the pre-shock region, and {\tt reflect} to account for X-ray reflection off the WD surface. 
Our physically motivated model then consists of {\tt 
 tbabs*pcfabs*(MCVSPEC*reflect+gauss)}. Furthermore, the {\tt constant} model was added to account for the flux variability between the different X-ray observations.  

\textcolor{black}{
We considered four representative accretion rates: the minimal value of $\dot{m} = 0.6$ g\,cm$^{-2}$\,s$^{-1}$ corresponding to $d = 3$ kpc and $f = f_{max}$ (Case A), two intermediate cases of $\dot{m} = 3 $ g\,cm$^{-2}$\,s$^{-1}$ (Case B) and $\dot{m} = 10 $ g\,cm$^{-2}$\,s$^{-1}$ (Case C), and the maximum value of  $\dot{m} = 44 $ g\,cm$^{-2}$\,s$^{-1}$ corresponding to  $d = 8$ kpc and $f = 0.001$ (Case D). We note that the shock height ($h_s$) is calculated and output from the {\tt MCVSPEC} model fit since it affects the reflection scaling factor ($r_{\rm{refl}}$) in the {\tt reflect} model.  For instance, in the tall accretion column, X-rays emitted near the shock are less likely to be reflected off the WD surface.  
Initially, we fit the model to the \xmm\ and \nustar\ spectra without a reflection ({\tt reflect}) component. This procedure is necessary to bound $h_s$ that can be used for calculating the reflection scaling factor ($r_{\rm{refl}}$), following the formula given by \cite{Tsujimoto2018}: $r_{\rm refl} = \Omega/2\pi = 1 - \sqrt{1-1/(1+h_s/R)^2}$. 
Since $h_s$ is calculated implicitly in the {\tt MCVSPEC} model using a shooting method and updating $v_{ff}$ iteratively in the accretion column profile computation, we cannot link $h_s$ and  $r_{\rm{refl}}$ directly in XSPEC. Thus, we need several iterations to fit the X-ray spectra by updating $h_s$ and adjusting $r_{\rm{refl}}$ in the reflection model. }
{\color{black} We then continued to fit the model, deriving new $M$ and $h_s$ values with each iteration and adjusting (and freezing) $r_{\rm{refl}}$ accordingly until $h_s$ did not change by more than $1 \%$ of the WD radius. In the end, we found that the reflection scaling factors  are 0.47, 0.76, 0.86, and 0.96 for the \mdot$ = 0.6$, \mdot$ = 3$, \mdot$ = 10$, and \mdot$ = 44$ g cm$^{-2}$ s$^{-1}$ cases, respectively. 
Note that a higher accretion rate reduces the shock height and thus increases the degree of X-ray reflection. 
}

\begin{figure}[ht]
\begin{minipage}{.5\textwidth}
 \hspace{2cm}
 \includegraphics[width=7cm]{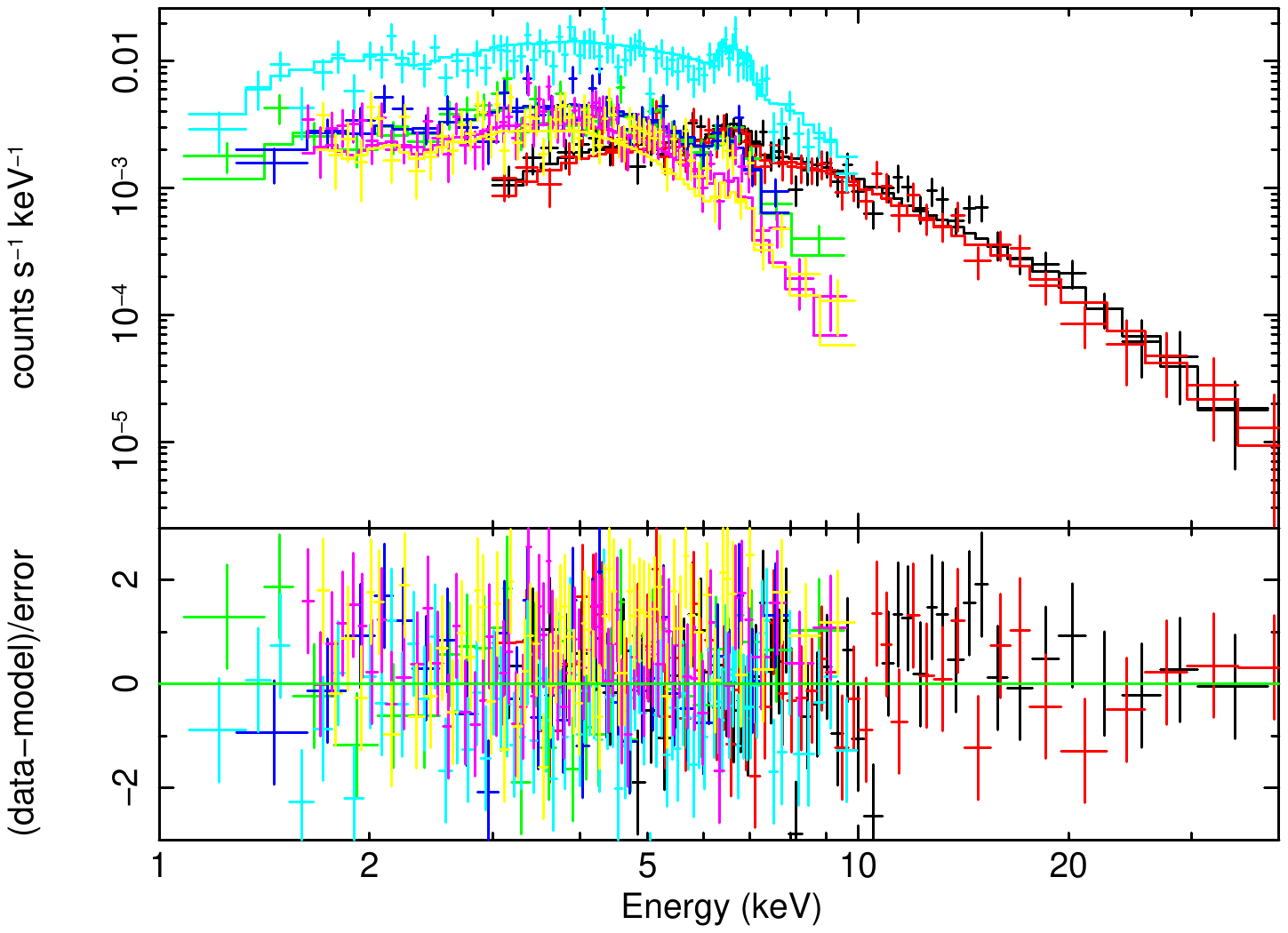}
\end{minipage}%
\begin{minipage}{.5\textwidth}
  \includegraphics[width=7cm]{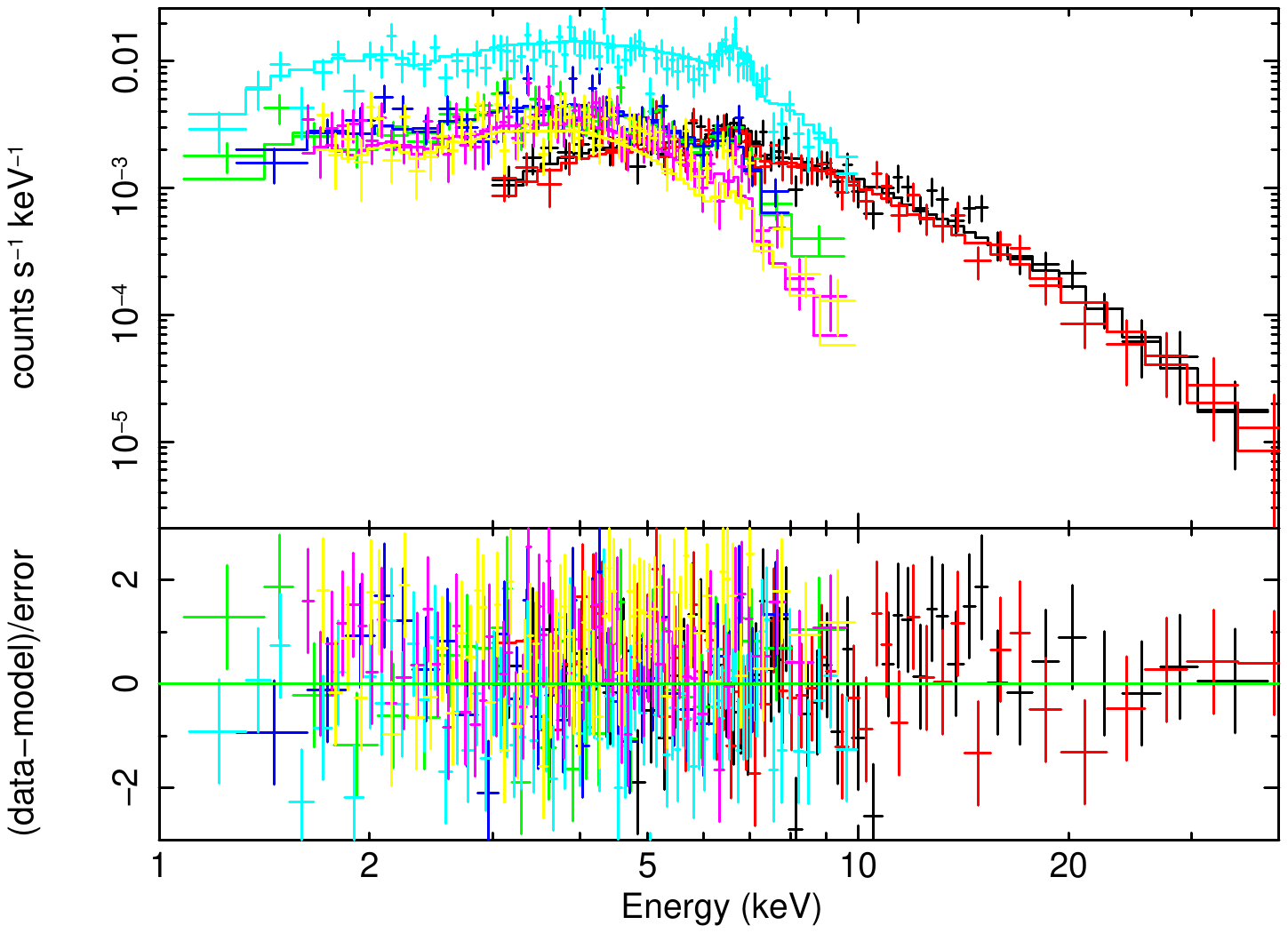}
\end{minipage}
\begin{minipage}{.5\textwidth}
 \hspace{2cm}
 \includegraphics[width=7cm]{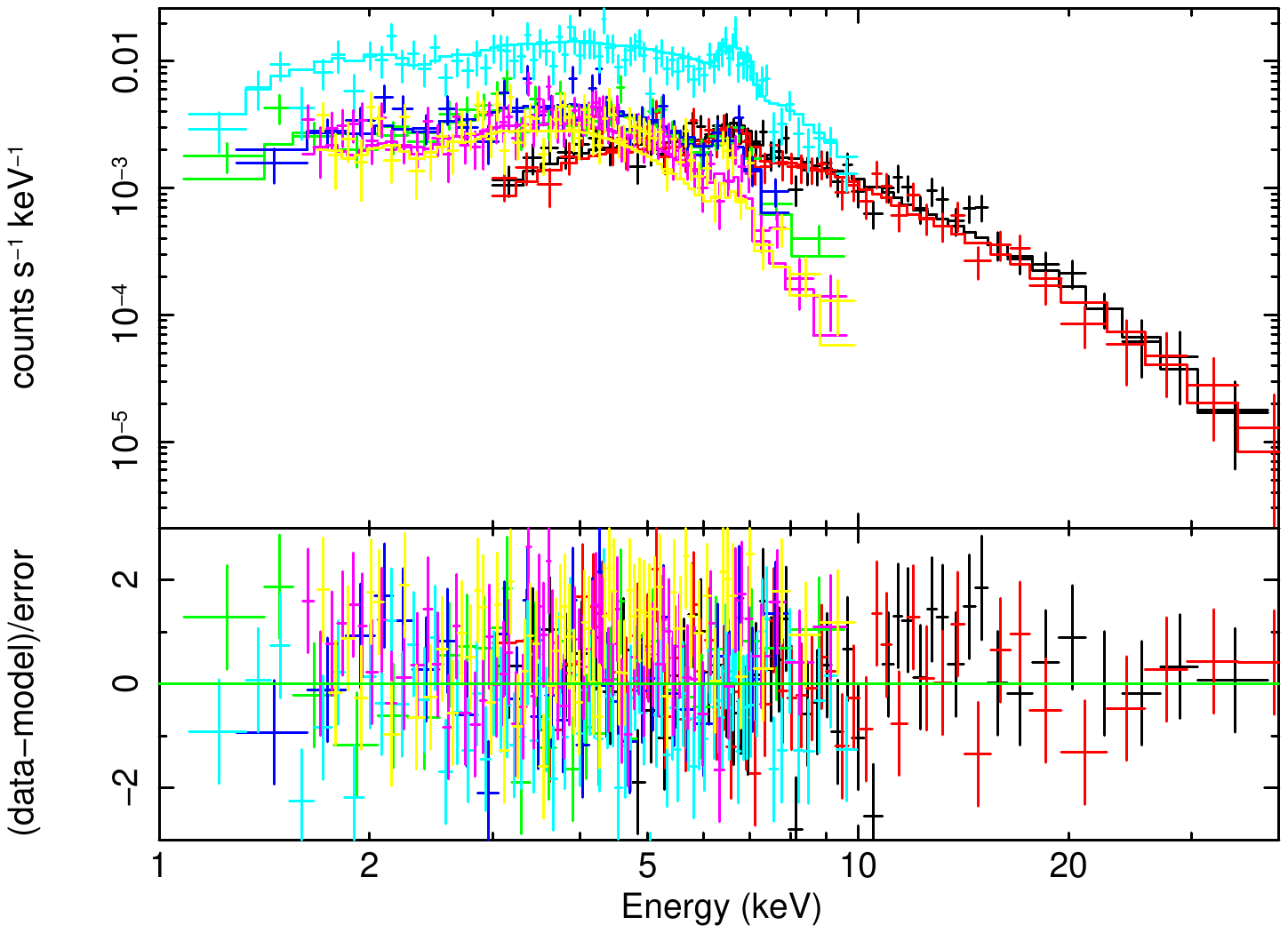}
\end{minipage}%
\begin{minipage}{.5\textwidth}
  \includegraphics[width=7cm]{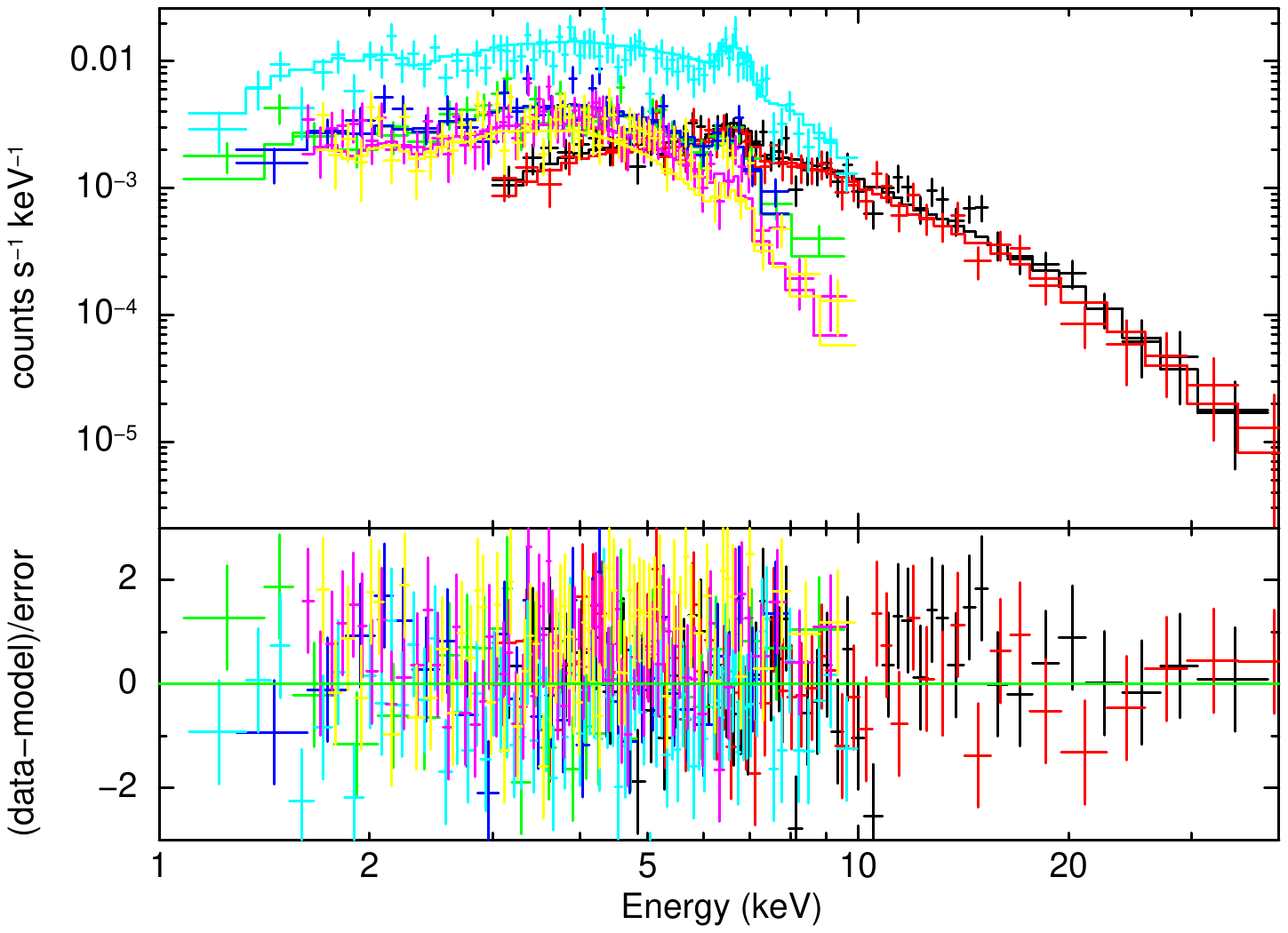}
\end{minipage}
\caption{\xmm\ and \nustar\ spectra fit by {\tt tbabs*pcfabs*(reflect*MCVSPEC+gauss)} and residuals for $R_{\rm M}/ R = 20$. Top Left: $\dot{m} = 0.6$ g\,cm$^{-2}$\,s$^{-1}$; Top Right: $\dot{m} = 3.0$ g\,cm$^{-2}$\,s$^{-1}$; Bottom Left: $\dot{m} = 10$ g\,cm$^{-2}$\,s$^{-1}$; Bottom Right: $\dot{m} = 44$ g\,cm$^{-2}$\,s$^{-1}$. }
\label{fig:mcvspec_fit} 
\end{figure}

{\color{black} Table \ref{tab:fits} presents our X-ray spectral fit results for the four accretion rates considered. For all the cases, the shock heights derived from the model remained below {\color{black}20 \%}  of the WD radius, ensuring the validity of the {\tt MCVSPEC} model, which assumes a planar geometry of the accretion column (compared to the dipole field configuration studied by \citet{Canalle2005} and \citep{Hayashi2014}). 
We found that all the cases yielded  similarly good fits to the \xmm\ and \nustar\ spectra with $\chi^2_\nu \sim 1.1$ (Figure \ref{fig:mcvspec_fit}). Fitting the atomic lines by {\tt MCVSPEC} also allows us to determine the Fe abundance to $Z = (0.8\rm{-}1.5) Z_\odot$.} 

\textcolor{black}{The best-fit WD mass depends on the assumed accretion rate more significantly for the lowest accretion rate (Case A) than in the other cases with higher $\dot{m}$ values (Case B--D) for two reasons. The low accretion rate case yields a much higher shock height of $h_s/R = 0.18$. First, it will reduce the free-fall velocity and make the model X-ray spectrum  softer  as a result of lower shock temperature, thus requiring to increase the WD mass to fit the X-ray spectral data. This effect was investigated by \citet{Hayashi2014} who found that mass accretion rates in the range of $\dot{m} \sim 1\rm{-}100$ [g\, cm$^{-2}$\,s$^{-1}$] did not alter the X-ray spectral shape. 
Secondly, the smaller shock height will increase the degree of X-ray reflection (i.e., larger $r_{\rm{refl}}$) and make the model X-ray spectrum harder \citep{Hailey2016}. Therefore, we need to decrease the WD mass in order to fit the X-ray spectra. Above $\dot{m} \sim 3.0 \;\mathrm{g\,cm^{-2}\,s^{-1}}$, the spectral parameters (including the WD mass) become insensitive to these two effects associated with $h_s$. This is because the small $h_s/R$ ratio no longer impacts the free-fall velocity, and also, the reflection fraction is close to unity. 
In this $\dot{m}$ range, the WD mass measurement converges to $\sim 0.81 \;M_\odot$ and is almost independent of the assumed  $\dot{m}$ value.}  

{\color{black} As described earlier, we should take into account the fact that the mass accretion rate is underestimated due to the (unobserved) cyclotron and soft X-ray blackbody emissions. First, we found that the cyclotron cooling effect is most pronounced for Case D when $\dot{m}$ is the largest. This is because, for the fixed $R_m/R = 20$, the WD B-field is largest at $B = 57$ MG as a result of the highest $\dot{m}$ value. From the best-fit WD mass and assumed $\dot{m}$ value, we calculated B-field using the formula for $R_m$ \citep{1989Norton}. However, even in this case with a higher B-field,  we found that $\epsilon_s = 0.5$ and $\bar{\epsilon} = 0.19$, indicating that a fraction of the cyclotron emission within the overall radiative cooling is 50\% and 19\% at the shock height and on average throughout the accretion column, respectively. Our results are consistent with the analytical solution of \citet{Wu1994} who found that  the X-ray spectral shape does not vary significantly due to the cyclotron cooling effect when $\epsilon_s < 20$  \citep{Wu1994}. 
To take into account the cyclotron emission in the bolometric luminosity, we increased the mass accretion rates by 5--19\% based on the derived $\bar{\epsilon}$ values. 
In addition, a fraction of the X-ray emission could be reprocessed through soft X-ray blackbody radiation. By taking the aforementioned IP 1RXSJ154814.5-452845 observed by \citet{Haberl2002} as a representative case, we estimate $\sim30$\% of the X-ray luminosity is missing through soft X-ray blackbody radiation since X-ray emission from our source is heavily absorbed below 1 keV. Overall, $\dot{m}$ may be underestimated by $\sim 35\rm{-}50$\% for Case A--D. Nevertheless, such increases in the $\dot{m}$ values do not change the WD mass significantly for Case B, C, and D because they are already in the high accretion regime where $M$ is insensitive to $\dot{m}$ variation as described above. For Case A where $\dot{m}$ increases from 0.6 to 0.8 g\,cm$^{-2}$\,s$^{-1}$, the WD mass goes down from $0.94\pm0.09 M_\odot$ to $0.92\pm0.08 M_{\odot}$. We adopt the latter WD mass value as a more realistic solution. However, we note that $\epsilon_s$ and $\bar{\epsilon}$ increase rapidly with the WD mass and exceed unity (meaning that the cyclotron cooling is more dominant over the thermal bremsstrahlung cooling), when it is above $\sim 1 M_\odot$. We caution that, for massive IPs with $M \simgt 1 M_\odot$ and without optical or UV detection, the ``missing" cyclotron cooling component should be estimated as shown above and reflected in the mass accretion rates.}

\begin{deluxetable*}{lcccccccc}[ht]
\tablecaption{{\tt MCVSPEC} fitting results}
\tablecolumns{5}
\tablehead{
\colhead{Parameter}   
& 
\colhead{Case A}
&
\colhead{Case B}
& 
\colhead{Case C}
& 
\colhead{Case D}
}
\startdata   
$R_m/R$ & 20 & 20 & 20 & 20\\ 
$\dot{m}^a$ [g\, cm$^{-2}$\,s$^{-1}$] & 0.6 &  3.0 &10 & 44\\ 
$M\, [M_{\odot}]$ & $0.94\pm0.09$ &  0.82$\pm 0.06$ &$0.81\pm{0.06}$  & $0.81_{-0.06}^{+0.07}$ \\	
$Z \, [Z_\odot]$ & $1.17_{-0.27}^{+0.25}$ &  1.03$_{-0.23}^{+0.29}$ & $1.02_{-0.23}^{+0.28}$ & $1.00_{-0.22}^{+0.26}$ \\	
$C_{2021}$ & $0.84\pm0.04$ &   0.85$\pm 0.04$ &$0.85_{-0.03}^{+0.04}$ & $0.86\pm0.04$ \\	
$C_{2010}$ & $1.13\pm0.04$ &  $0.97\pm0.04$ &$0.97_{-0.03}^{+0.04}$&  $0.97\pm0.04$\\	
$N^{(i)}_H (10^{22} \rm{cm}^{-2})$ & $2.5\pm0.5$ & 2.5$\pm0.5$ & $2.5\pm0.5$ &  $2.5\pm0.05$\\	
$N^{(p)}_H (10^{22} \rm{cm}^{-2})$ & $11.2_{-1.4}^{+1.7}$ & 11.0$_{-1.3}^{+1.6}$ &$11.0_{-1.27}^{+1.57}$ & $10.9_{-1.3}^{+1.6}$ \\	
$c^b_f$ & $0.79_{-0.06}^{+0.05}$ & $0.80\pm0.05$ & $0.80\pm0.05$& $0.80\pm0.05$\\	
$r_{\rm refl}$ & $0.47$ & 0.76 & $0.86$ & $0.96$\\	
$\cos(i)$ & $0.95_{-0.39}^{+0.00}$ & $0.95_{-0.36}^{+0.00}$ & $0.95_{-0.36}^{+0.00}$ & $0.95_{-0.36}^{+0.00}$\\	
$\chi^2_{\nu}$ (dof) & 1.12 (371) & 1.11 (371) & 1.11 (371) & 1.11 (371) \\	
$h_s/R$ & $18 \%$ & $3 \%$ & $1 \%$ &  $< 1 \%$\\	
$\epsilon_s ^{~c}$ & 0.11 & 0.19 &0.24 & 0.50 \\
$\bar{\epsilon} ^{~d}$ & 0.05 & 0.07 & 0.10 & 0.19 \\
B (MG)$^e $ & 7 & 15 & 27 & 57 \\
\enddata
\label{tab:fits}
    All errors shown are 1-$\sigma$ intervals. \\
    $^a$ Specific accretion rate is presented without error bars as it was frozen for fits.\\
    $^b$  $c_f$ is the covering fraction associated with {\tt pcfabs}, not to be confused with the fractional accretion area.\\
    $^c$ The ratio of the bremmstrahlung cooling timescale to cyclotron cooling timescale at the shock (see \cite{Wu1994}). This is derived from the fit parameters. \\
    $^d$ The mean ratio of the bremmstrahlung cooling timescale to cyclotron cooling timescale along the accretion column. This is derived from the profiles of the model. \\
    $^e$ An estimate of magnetic field strength derived from the fit parameters.\\
\end{deluxetable*}

\subsubsection{IP parameter determination} 

{\color{black} While the statistical errors for $M$ are small ($<10$\%) due to the high-quality X-ray data, the systematic errors arising from the unknown mass accretion rate are more predominant as manifested by the discrepancy between Case A ($M = 0.94 M_\odot$) and the other cases ($M = (0.81\rm{-}0.82)M_\odot$).  As discussed earlier, $\dot{m}$ depends on both the source distance and fractional accretion column area. 
Therefore, if we take the broadest parameter space for $d$ and $f$, the WD mass was constrained to $M = (0.75\rm{-}1.03)M_\odot$. If we assume that the source is in the Galactic Center (at $d = 8$ kpc), the minimum possible accretion rate would be $\dot{m} = 1.5 \;\mathrm{g\,cm^{-2}\,s^{-1}}$, corresponding to an $h_s/R \sim 7 \%$ and $r_{\rm{refl}} = 0.64$. In this case, $M = (0.85\pm0.06)M_\odot$, and thus the WD mass is in the range  of $M = (0.75\rm{-}0.91) M_\odot$, consistent with the mean WD mass of IPs \citep{Zorotovic2011, Shaw2020}.  

Based on the best-fit WD masses, $R_m/R = 20$ and assumed mass accretion rates, we estimated the WD B-field using the magnetic radius formula \citep{Norton2004} -- $B = 7, 15, 27$ and 57 MG for Case A, B, C and D, respectively. The loose constraints on the WD B-field in the range of $B = 7\rm{-}57$ MG are on the higher end compared to the typical range for IPs ($B\sim0.1\rm{-}10$ MG). The B-fields in Case C and D are comparable to or even exceed the highest IP B-field ever measured from  V405 Aur  \citep[$B= 32$ MG; ][]{Piirola2008}.

}

\section{Conclusion}
{\color{black}
    We have both provided further analysis on J1745 and presented a novel methodology for constraining the IP parameters, including the WD mass and B-field. We presented further evidence, using X-rays below $\sim$ 5 keV, that there is a statistically significant 614-second period. With our enlarged \xmm\ data set, we do not confirm the asymmetry in the pulse shape of the 1227-second X-ray periodicity seen by Gong (2022). Consequently, we argue that the 614-second period is the spin period, the 1227-second period is a lower harmonic, and the emission is from one pole only. Using those results and a wide range of accretion rates and our novel X-ray spectroscopic model {\tt MCVSPEC}, we measured the WD mass of J1745. Furthermore, our physically-motivated model encapsulates many features and corrections that are not found together in other IP models, such as the effect of shock height and magnetosphere radius. Its calculation of shock height also serves as a means to constrain the effects of X-ray reflection of the WD surface of an IP. As demonstrated in our study of J1745-3213, it is essential to follow up X-ray sources (which may show periodic signals and/or large hardness ratios as potential signatures of magnetic CVs) with \nustar\ for firmly determining the source types and WD masses. Our results demonstrate promising outcomes from mCV populations in the X-ray band as a complementary approach to the optical/UV studies \citep{Pala2022}. 
    
    Furthermore, our methodology may be applied to its greatest effect on sources of known or well-constrained distance. This paper presents the first application of a comprehensive X-ray spectral model to a newly discovered IP, incorporating the finite magnetosphere radius, the impact of shock height, and X-ray reflection in a self-consistent manner. By accounting for uncertainties associated with  source distance and fractional accretion column area, we have determined the systematic errors related to the WD mass. Our methodology could be applied to other IPs for measuring their WD masses more accurately. Notably, recent dynamical measurements of WD masses in GK Per and XY Ari have presented challenges to the previous (and differing) WD mass measurements using X-ray spectroscopy data \citep{Alvarez2021, Alvarez2023}. To reduce systematic errors, we stress the significance of measuring source distances, detecting or estimating cyclotron emission in the optical/UV band (especially for highly-magnetized IPs), and obtaining broad-band X-ray spectra beyond 10 keV. 
    In the future, the HEX-P X-ray probe mission, which covers a broad X-ray band of 0.1--80 keV, will be ideal for determining the WD masses accurately from a substantial number of mCVs \citep{Madsen2019HEX}. }

\section{Acknowledgements}
 When we were preparing our manuscript for re-submission, we became aware of \citet{Gong2022} who analyzed the same \xmm\ and \nustar\ data of \src. We thank the referee for carefully reading our manuscript and making valuable comments. We thank Dr. Matteo Bachetti for helping us with the X-ray timing analysis. 
KM acknowledges support from \nustar\ Cycle 6 Guest Observer Program grant NNH19ZDA001N. 
GP acknowledges funding from the European Research Council (ERC) under the European Union’s Horizon 2020 research and innovation program (grant agreement No 865637). CGS acknowledges continuous support from Cristina "Yaya" Salcedo during this project and the development of {\tt MCVSPEC}.
\software{HEASoft Version 6.25 \citep{heasoft}, FTools Version 6.25 \citep{heasoft}}

\bibliography{main} 

\appendix{
\vspace{-.7cm}
\subsection{\xmm-EPIC folded lightcurves}
We present phase profiles folded with 614-sec, 1228-sec, and 1841-sec periods with the newest and highest quality available data to justify that the 614-sec signal represents a fundamental period. The symmetry between two peaks that appear in a single phase profile indicates each peak is the same signal repeated twice. All higher harmonics of the 614-sec periodicity show peaks within error of each other. Furthermore, the dip between the peaks goes below the mean  counts (red horizontal lines), indicating that these are non-overlapping signals. It must be noted, however, that the original observation was excluded as it anomalously only detected our source with the MOS1 instrument. 
In summary, based on the new \xmm\ observation of the source on-axis, for which we provide a joint pulse profile from all three modules, indicates the signal at 1228 sec is a repeated 614-sec signal.
\begin{figure}[h]
\hspace{-1cm}                                                       
\begin{minipage}{.1\textwidth} 
  \includegraphics[width=19.5cm]{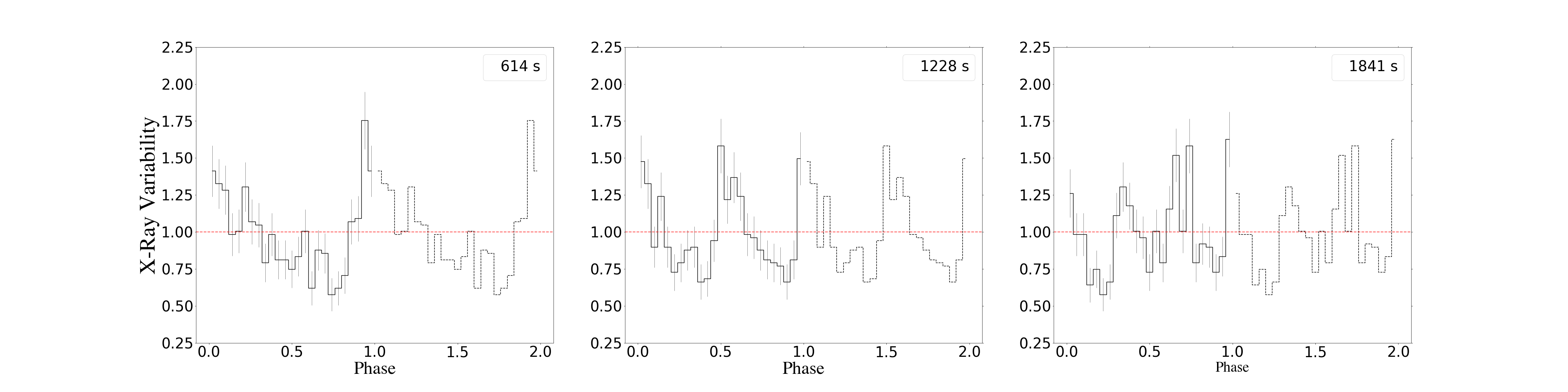}
  \label{fig:newMOSphase1}
\end{minipage}

\hspace{-1cm}                                                       
\begin{minipage}{.1\textwidth}
  \includegraphics[width=19.5cm]{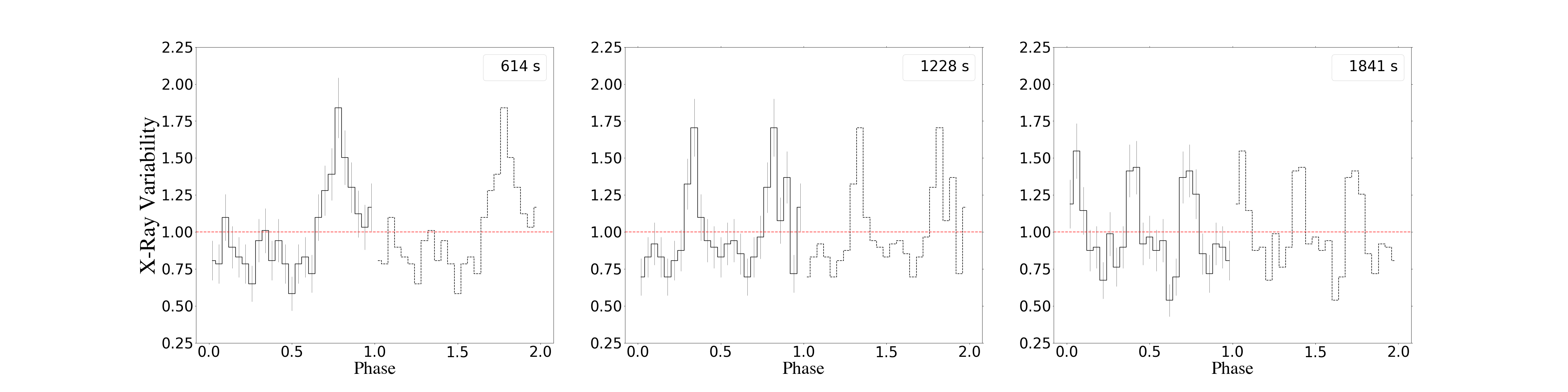}
  \label{fig:newPNphase1}
\end{minipage}

\hspace{-1cm}                                                       
\begin{minipage}{.1\textwidth}
  \includegraphics[width=19.5cm]{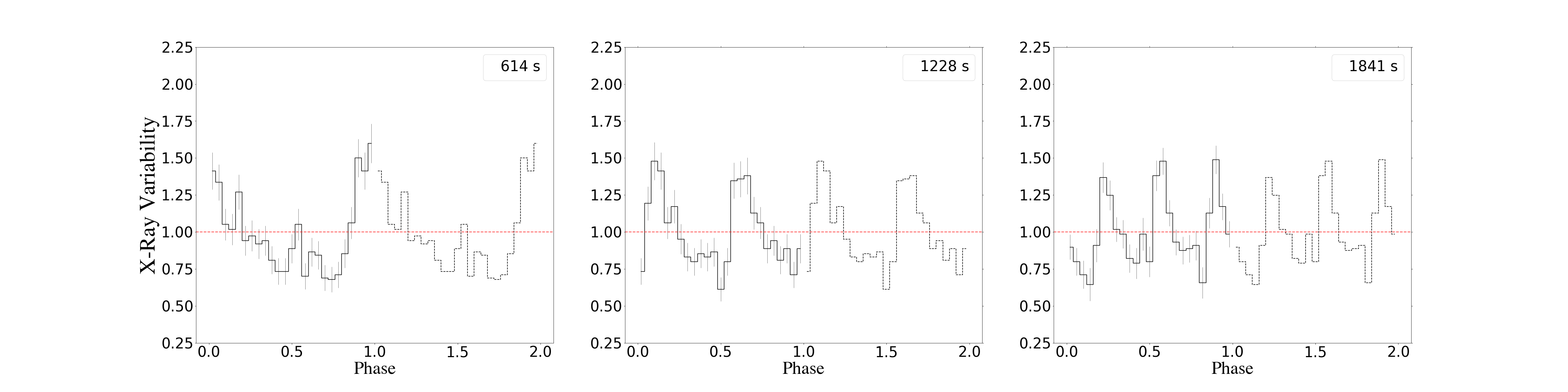}
  \label{fig:jointphase1}
\end{minipage}

  \caption{ Pulse profiles at the peak frequency and its harmonics for 0870990201 MOS (top), 0870990201 PN (middle), 0870990201 joint (bottom). The red horizontal lines indicate the mean counts per phase bin and 1-$\sigma$ error bars are included at  $\phi=0-1$.}

\end{figure}

\subsection{Methodology for determining WD mass and magnetic field strength}
\begin{figure}[ht]
\includegraphics[width=18cm]{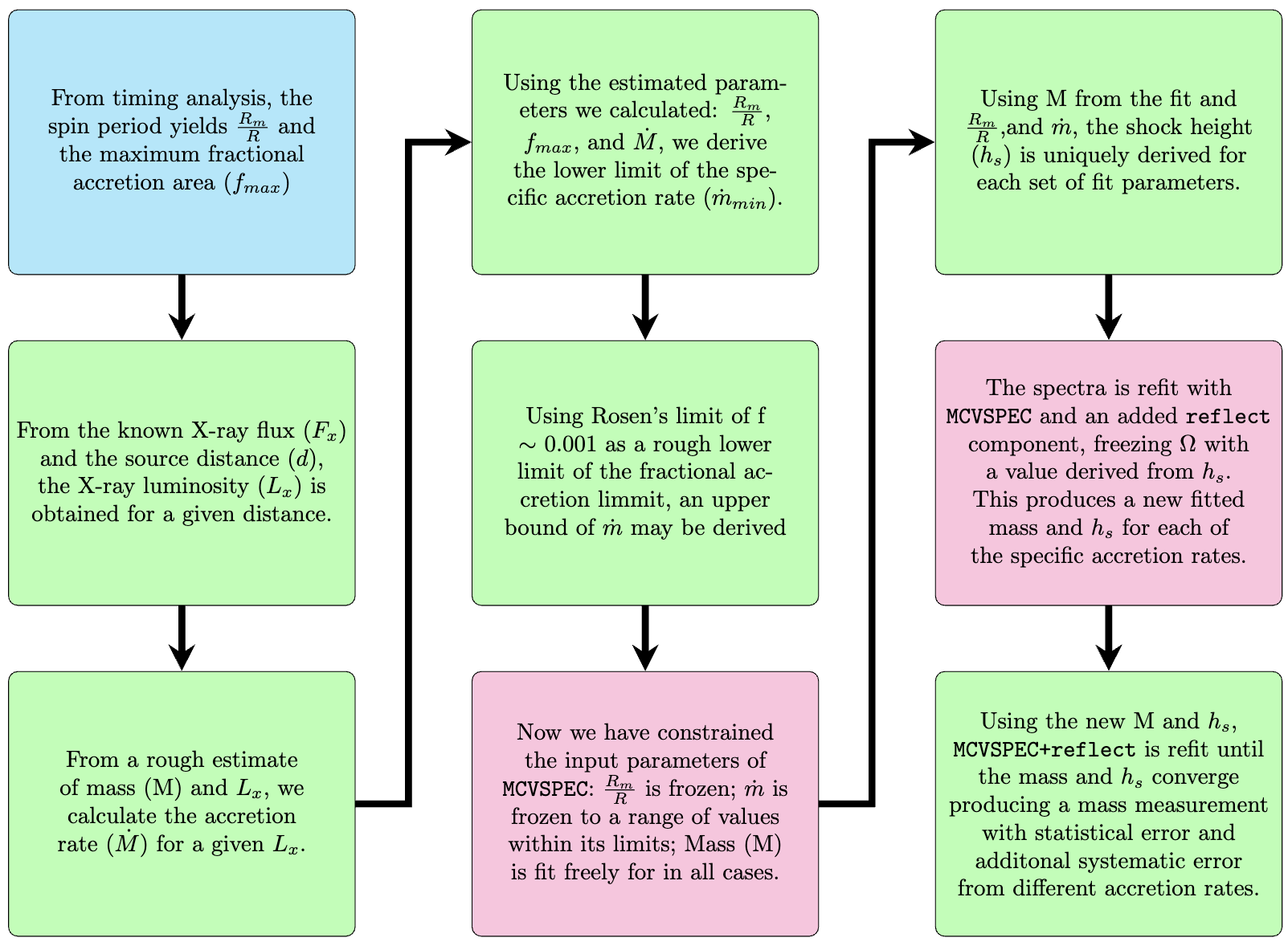}

  \caption{The flowchart demonstrates the procedures for determining WD $M$. The blue box indicates the step related to X-ray timing analysis. We derived variables from the known or fit parameters in the steps colored in green. Magenta boxes indicate the steps involved in spectral fittings.}   
\end{figure}
}

\end{document}